\documentclass[groupedaddress,
twocolumn,
showkeys,
  preprintnumbers,
amsmath,amssymb,
  aps,
  pre,
floatfix,
  ]{revtex4-1}

\usepackage{dynlearn}
\usepackage{amsthm}
\usepackage{mathrsfs}
\usepackage{graphicx}\usepackage{dcolumn}\usepackage{bm}

\theoremstyle{definition}

\renewcommand{\H}{\operatorname{H}}

\newcommand{\kB} { k_\text{B} }
\newcommand{\cs}{\causalstate}

\newcommand{\One}{ {\mathbf{1} } }
\newcommand{\MSym}{\MeasSymbol}
\newcommand{\msym}{\meassymbol}

\newcommand{\Abet}{\ProcessAlphabet}
\newcommand{\SSet}{\CausalStateSet}

\newcommand{\MxSSet}{\AlternateStateSet}
\newcommand{\MxSMeasure}{\mu}
\newcommand{\MxSDyn}{\mathcal{W}}
\newcommand{\mxst}{\eta}

\newcommand{\TapeVar}{X}
\newcommand{\TapeSym}{x}
\newcommand{\TapeAlphabet}{\mathcal{X}}
\newcommand{\RatchetStateVar}{R}
\newcommand{\RatchetStateSym}{r}
\newcommand{\RatchetStateAlphabet}{\mathcal{R}}

\newcommand{\RatchetMC}{M}
\newcommand{\InputStateVar}{S}
\newcommand{\InputStateSet}{\mathcal{S}}

\newcommand{\CmuDim}{d_{\mu}}
\newcommand{\LCEDim}{d_\text{LCE}}

\newcommand{\WeightParameter}{\epsilon}
\newcommand{\CoarseGrain}{\varepsilon}

\begin{document}

\title{Functional Thermodynamics of Maxwellian Ratchets:\\ 
Constructing and Deconstructing Patterns,\\
Randomizing and Derandomizing Behaviors}

\author{Alexandra M. Jurgens}
\email{amjurgens@ucdavis.edu}

\author{James P. Crutchfield}
\email{chaos@ucdavis.edu}

\affiliation{Complexity Sciences Center and Physics Department,
University of California, Davis, California 95616}

\date{\today}
\bibliographystyle{unsrt}

\begin{abstract}

Maxwellian ratchets are autonomous, finite-state thermodynamic engines that
implement input-output informational transformations. Previous studies of these
``demons'' focused on how they exploit environmental resources to generate
work: They randomize ordered inputs, leveraging increased Shannon entropy to
transfer energy from a thermal reservoir to a work reservoir while respecting
both Liouvillian state-space dynamics and the Second Law. However, to date,
correctly determining such functional thermodynamic operating regimes was
restricted to a very few engines for which correlations among their
information-bearing degrees of freedom could be calculated exactly and in
closed form---a highly restricted set. Additionally, a key second dimension of
ratchet behavior was largely ignored---ratchets do not merely change the
randomness of environmental inputs, their operation constructs and deconstructs
patterns. To address both dimensions, we adapt recent results from
dynamical-systems and ergodic theories that efficiently and accurately
calculate the entropy rates and the rate of statistical complexity divergence
of general hidden Markov processes. In concert with the Information Processing
Second Law, these methods accurately determine thermodynamic operating regimes
for finite-state Maxwellian demons with arbitrary numbers of states and
transitions. In addition, they facilitate analyzing structure versus randomness
trade-offs that a given engine makes. The result is a greatly enhanced
perspective on the information processing capabilities of information engines.
As an application, we give a thorough-going analysis of the Mandal-Jarzynski
ratchet, demonstrating that it has an uncountably-infinite effective state
space.

\end{abstract}

\keywords{Nonequilibrium thermodynamics, Information Processing Second
Law, Kolmogorov-Sinai entropy, Shannon entropy rate, causal states, mixed
states, ergodicity, contraction maps, place-dependent iterated function systems}
\maketitle

\preprint{\arxiv{2003.00139}}

\section{Introduction}
\label{sec:Introduction}

In 1867, James Clerk Maxwell introduced a thought experiment designed to
challenge the Second Law of Thermodynamics \cite{Maxw88a, Maruyama09}; what Lord
Kelvin later came to call ``Maxwell's Demon''. Exploiting the fact that the
Second Law holds only on average---i.e., the thermodynamic entropy $\langle S
\rangle $ cannot decrease over repeated transformations---the experiment
conjured an imaginary, intelligent being capable of detecting and then
harvesting negative entropy fluctuations to do work. The paradox that Maxwell
put forward is that by using its ``intelligence'' this being apparently
violates the Second Law of Thermodynamics. Maxwell's challenge was the first
indication that the Second Law must take into account information processing.
A century and a half later, many now appreciate that this is critical to future
progress in the molecular and nanoscale sciences and engineering.

The puzzle's solution came from recognizing that the ``very observant'' and
``neat-fingered'' Demon must manipulate memory to perform its detection and
control task and, critically, that such information processing comes at a cost
\cite{Penrose70, Bennet82}. To operate, the Demon's intelligence has
thermodynamic consequences. This is summarized by Landauer's Principle: ``any
logically irreversible manipulation of information ... must be accompanied by a
corresponding entropy increase in non-information-bearing degrees of freedom of
the information-processing apparatus or its environment'' \cite{Land61a}. This
recasts the Demon as a type of engine---an \emph{information engine} that uses
correlations in an \emph{information reservoir} to leverage thermodynamic
fluctuations in a \emph{heat reservoir} to do useful work.

This class of information engines---Maxwellian demons and their generalized
ratchets---has been subject to extensive study \cite{Mand012a, Boyd15a,
Boyd16c, Boyd16e}. However, previous determinations of their
thermodynamic functionality were stymied by the difficulty of accurately
calculating the entropic change in what Landauer identified as the system's
``information-bearing degrees of freedom''.

Consider a Maxwellian ratchet designed to read an infinite input tape, perform a
computation and thermodynamic transformation, and write to an infinite output
tape, as depicted in \cref{fig:InfoRatchet}. The relevant entropic change then
is quantified by the difference in the Kolmogorov-Sinai entropies of the inputs
to the ratchet ($\hmu$) and  of the outputs to the information reservoir
($\hmu^\prime$) \cite{Boyd15a}. However, in general, this calculation ranges
from very difficult to intractable when the processes generating the input and
output information have temporal correlations. And, more troubling, this problem
is generic in the space of finite-state ratchets. When driving a ratchet with
uncorrelated input, even simple memoryful ratchets produce output processes with
temporal correlations. Fundamental progress was halted since determining
thermodynamic functionality in the most general case---temporally correlated
input driving a memoryful ratchet---was intractable. Attempts to circumvent
these problems either heavily restricted thermodynamic-controller architecture
\cite{Boyd15a}, invoked approximations that misclassified thermodynamic
functioning, or flatly violated the Second Law \cite{Mand012a}. It appears
that---and this is one practical consequence of the results reported in the
following---a number of recent analyses of information-engine efficiency and
functioning must be revisited and corrected. Our contribution is that the
latter is now possible.

Re-examining a well-known information ratchet, we introduce techniques to
accurately measure the Kolmogorov-Sinai-Shannon entropy of temporally correlated
processes in general. We show that, via the Information Processing Second Law
\cite{Boyd15a}, this allows accurately determining the functional thermodynamics
of arbitrary finite-state ratchets. Notably, the net result is a shift in
perspective. To guarantee that the output information could be studied
analytically, previous successful efforts designed ratchet
\emph{structure}---the states and transitions---in accord with a given input's
correlational structure \cite{Boyd16c}. One consequence is that follow-on
efforts adopted a fixed input-output-centric view of information engines. Here,
following the example of the earliest discussions of information ratchets
\cite{Mand012a}, the new methods shift the focus back to the engine itself,
setting its design and then exploring all possible input-dependent thermodynamic
functionalities.

The approach has appeal beyond mere narrative and historical symmetry. The shift
in focus reveals a second dimension to ratchet functionality. The change in
entropy rate $\Delta \hmu = \hmu^\prime - \hmu$ monitors the degree to which the
ratchet transforms a process' informational content, but it does not address
\emph{how} this comes about. To do this requires investigating the change in
structure from the input process to the output process. These structural changes
were previously proved to be deeply relevant to engine thermodynamic efficiency
and an engine's ability to meet the work production bounds set by Landauer's
Principle \cite{Boyd17a}. Their impact is nontrivial. For example, we will show
that forcing a Maxwellian ratchet to perfectly generate or erase structure
requires divergent memory resources. 

To reach this conclusion, the next section briefly reviews information engines
and ratchets, including their energetics, structure, and informatics. This,
then, allows us to highlight the calculational intractability for generic
thermodynamic ratchets. To be concrete, we recall one of the first ratchets and
review how its thermodynamic functionality---engine, eraser, or dud---is
determined. Using new methods from ergodic theory and dynamical systems that
determine randomness generation and memory use (recounted in the Supplementary
Materials), we then re-analyze the original ratchet, showing that previous
analyses misidentified its thermodynamic functioning. This is illustrated for
its operation in several distinctly-correlated environments. We then explore
the structural dimension of ratchet functionality, demonstrating that the
engine/eraser/dud classification does not uniquely describe ratchet information
processing for a given input. To remedy this, in conjunction with the previous
functional classification which can now be exactly carried out, we introduce
\emph{structure-randomness} trade-offs in engine operation, highlighting the
multi-dimensional nature of ratchet information processing.

\section{Information Engines}
\label{sec:InfoEngines}

\begin{figure}
\centering
\includegraphics[width=0.47\textwidth]{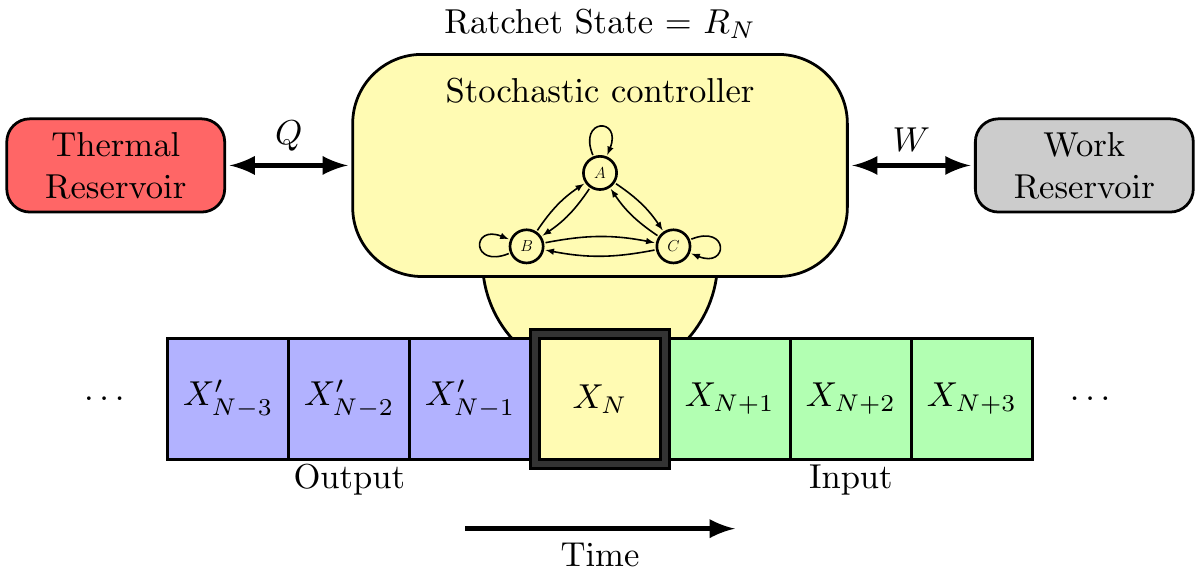}
\caption{Information engine as a finite-state ratchet (controller) connected to
	a thermal reservoir, a work reservoir, and an information reservoir
	(depicted as tape whose storage cells may be read or written).
	}
\label{fig:InfoRatchet}
\end{figure}

The information engines of interest consist of a finite-state stochastic
controller or \emph{ratchet} that interacts with a \emph{thermal reservoir}, a
\emph{work reservoir}, and an \emph{information reservoir}. These are connected
as shown in \cref{fig:InfoRatchet} and are embedded in a thermal environment at
constant temperature $T$. The information reservoir takes the form of an
\emph{input tape}, which stores a binary-symbol string. Its state is described
by the random variable $\TapeVar_{0:\infty} = \TapeVar_0 \TapeVar_1 \dots$. We
restrict to binary input and output alphabets, so that each $\TapeVar_N$
realizes an element $\TapeSym_n \in \TapeAlphabet = \{0, 1\}$. The ratchet
operates in continuous time; the controller state at time $t = N \tau$ is
represented by the random variable $\RatchetStateVar_N$, which realizes an
element $\RatchetStateSym \in \RatchetStateAlphabet$---the ratchet's discrete,
finite state space.

At each step, $\TapeVar_N$ couples to the ratchet controller for an interaction
of duration $\tau$. During this time, thermal fluctuations continuously drive
transitions in the coupled state space $\RatchetStateAlphabet \times
\TapeAlphabet$ of the ratchet and the current tape symbol. After the
interaction interval, the ratchet is in a potentially different state
$\RatchetStateVar_{N+1}$, and the symbol $X_N$ has been transduced into an
output symbol $\TapeVar'_N = \TapeSym'_N \in \TapeAlphabet$, which is written
to the tape. The strings of possible output symbols are expressed by the random
variable $\TapeVar'_{0:\infty} = \TapeVar'_0 \TapeVar'_1 \dots$. The tape moves
forward, and the next input symbol $\TapeVar_{N+1}$ begins its interaction with
the ratchet, which now starts in state $\RatchetStateVar_{N+1}$. The joint
transitions between states of the ratchet and bit have energetic consequences,
capturing energy flows between the thermal and work reservoirs.

\subsection{Energetics}
\label{sec:InformationEngineEnergetics}

These information engines are autonomous and transitions in the coupled
ratchet-symbol system are driven by fluctuations in the thermal reservoir.
Recently, Ref. \cite{Boyd16c} introduced a general formalism for determining the
energetics of such information engines. First, noting that the ratchet-symbol
system obeys detailed balance, transitions over the joint ratchet-symbol state
space $\RatchetStateAlphabet \times \TapeAlphabet$ is described by a Markov
chain $\RatchetMC$, where every transition with positive probability---denoted:
\begin{align*}
  & M_{\RatchetStateSym_N \otimes \TapeSym_N \to \RatchetStateSym_{N+1} \otimes \TapeSym'_N} = \\
  & \:\:\: \Pr(\RatchetStateVar_{N+1} = \RatchetStateSym_{N+1}, \TapeVar'_N = \TapeSym'_N | \RatchetStateVar_N = \RatchetStateSym_N, \TapeVar_N = \TapeSym_N )
\end{align*}
---must have a reverse transition with positive probability. Energy changes
associated with an internal-state transition are then determined by the
forward-reverse transition probability ratio:
\begin{align*}
  \Delta E_{\RatchetStateSym_N \otimes \TapeSym_N \to \RatchetStateSym_{N+1}
  \otimes \TapeSym'_N}
  = \kB T \ln \frac{M_{\RatchetStateSym_{N+1} \otimes
  \TapeSym'_N \to \RatchetStateSym_N \otimes \TapeSym_N}}
  {M_{\RatchetStateSym_N \otimes \TapeSym_N \to \RatchetStateSym_{N+1} \otimes \TapeSym'_N}}
  ~.
\end{align*}
Assuming that all energy exchanges with the heat reservoir occur during the
ratchet-symbol interaction interval $\tau$ and that all energy exchanges with
the work reservoir occur between interaction intervals, the average asymptotic
work is:
\begin{align}
  \langle W \rangle
  = \sum_{\RatchetStateSym, \RatchetStateSym' \in \RatchetStateAlphabet,
  \TapeSym, \TapeSym' \in \TapeAlphabet}
  \pi_{\RatchetStateSym \otimes \TapeSym}
  M_{\RatchetStateSym \otimes \TapeSym \to \RatchetStateSym' \otimes \TapeSym'}
  \Delta E_{\RatchetStateSym \otimes \TapeSym \to \RatchetStateSym' \otimes \TapeSym'}
  ~,
  \label{eq:WorkProduction}
\end{align}
where $\pi_{\RatchetStateSym \otimes \TapeSym}$ is the asymptotic distribution
over the joint state of the ratchet-symbol system at the beginning of an
interaction interval.

\subsection{Structure}
\label{sec:InformationEngineStructure}

To discuss the computational \emph{structure} of information engines, we first
cast the input and output strings in terms of the hidden Markov Models (HMMs)
that generate them.

\begin{Def}
\label{Def:HMM}
A finite-state edge-labeled \emph{hidden Markov model} (HMM) consists of:
\begin{enumerate}
\setlength{\topsep}{0mm}
\setlength{\itemsep}{0mm}
\item A finite set of states
	$\CausalStateSet = \{\causalstate_1, ... , \causalstate_N \}$,
\item A finite alphabet $\MeasAlphabet$ of $k$ symbols
	$\msym \in \MeasAlphabet$, and
\item A set of $N$ by $N$ symbol-labeled transition matrices
    $T^{(\msym)}$, $\msym \in \MeasAlphabet$:
    $T^{(\msym)}_{ij} = \Pr(\causalstate_j,\msym|\causalstate_i)$.
    The corresponding overall state-to-state transitions are described by the
	row-stochastic  matrix $T = \sum_{\msym \in \MeasAlphabet} T^{(\msym)}$.
\end{enumerate}
\end{Def}

This information-ratchet representation allows us to consider the internal
states $\InputStateSet$ of the \emph{input machine} (HMM) as well as the
internal states $\InputStateSet'$ of the \emph{output machine} (HMM). The
latter are the joint states of the input process and the ratchet:
$\InputStateSet' = \InputStateSet \times \RatchetStateAlphabet$.

When the string of inputs or outputs can be generated by an HMM with only a
single internal state, they are \emph{memoryless}, since they can store no
information from the past. The random variables generated by the associated HMM
are \emph{independent and identically distributed} (IID). When there is more
than a single state, in contrast, the associated process is \emph{memoryful}
and the random variables generated may be correlated in time.

\begin{figure*}
\centering
\includegraphics[width=\textwidth]{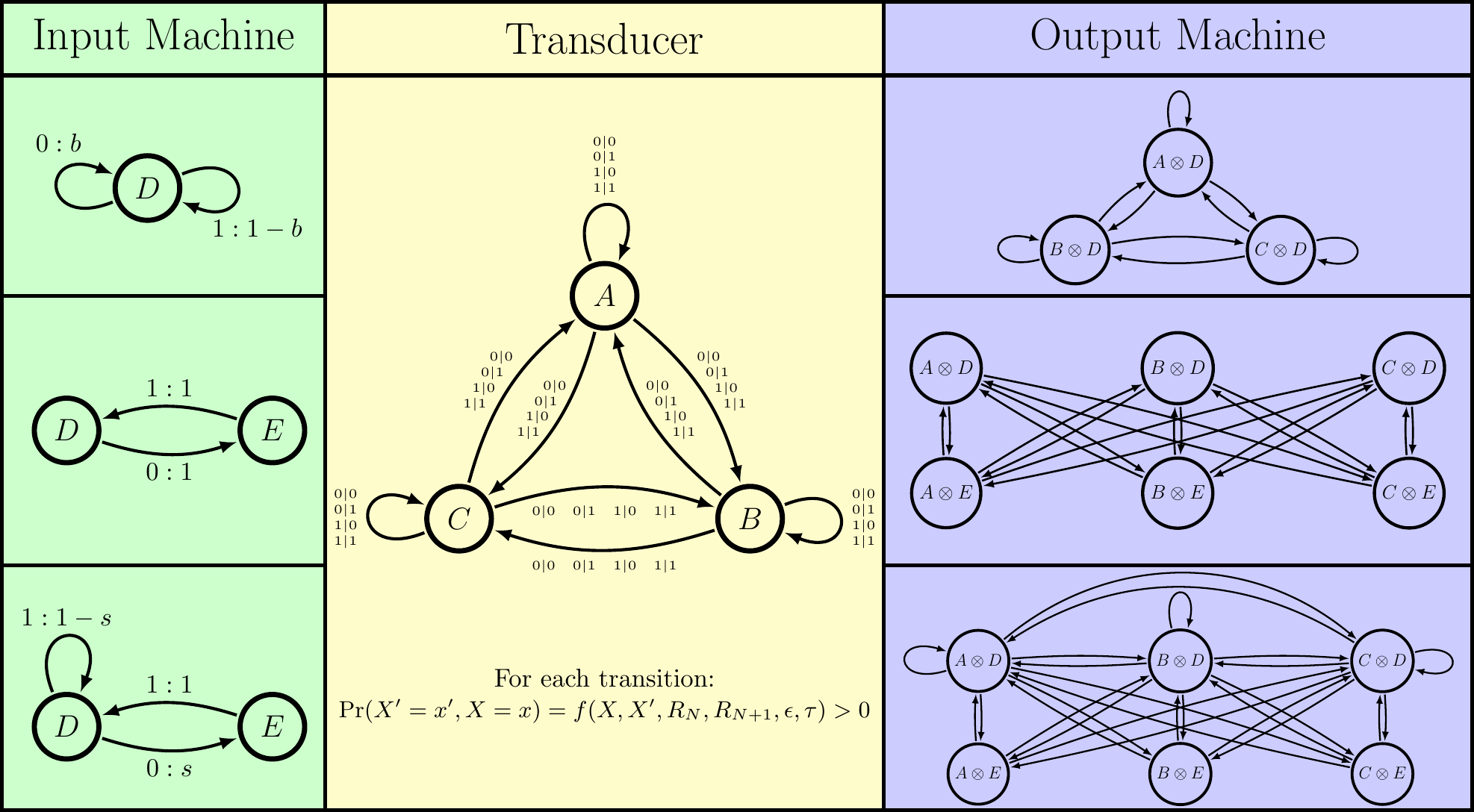}
\caption{Composing the Mandal-Jarzynski transducer (center, yellow) with a
	Hidden Markov model (left, green) that describes the process on the input
	tape gives an output Hidden Markov model (right, purple) that describes the
	process written to the output tape. Hidden Markov model (HMM) states
	$\RatchetStateVar$ are depicted as circles. Directed edges between states
	represent possible transitions on an observed symbol $\msym$. HMM edges are
	labeled $\msym : \Pr(\msym', \RatchetStateVar_{N+1} \vert
	\RatchetStateVar_N)$. Transducers are similarly depicted by circular states
	with directed edges representing possible transitions on pairs of input
	symbols $\msym$ and output symbols $\msym'$.  Transducer transitions are
	specified by $\msym' \vert \msym : \Pr(\msym', \RatchetStateVar_{N+1} \vert
	\msym', \RatchetStateVar_N ) $. Left: Input HMMs discussed here, from top
	to bottom, a (memoryless) Biased Coin, a Period-2 Process, and the Golden
	Mean Process. Center: The Mandal-Jarzynski ratchet, represented by a
	three-state transducer. Probabilities are not shown on edge labels for
	conciseness, but are nonzero for all transitions and all combinations of
	input-output symbol pairs $(\msym, \msym')$. Each edge probability is a
	function---denoted by $f(\ldots)$---of the previous state
	$\RatchetStateVar_N$, the next state $\RatchetStateVar_{N+1}$, the input
	symbol $\msym$, and the output symbol $\msym'$. See \cref{sec:MJRatchet}
	and \cref{app:MandalJarzynski} for further details. Right: Output HMMs
	resulting in the Mandal-Jarzynski transducer composed with the
	corresponding input HMM on left. Edge labels are left off for conciseness,
	but each transition label represents a positive probability of observing a
	$0$ or a $1$.
	}
\label{fig:MJ_comp_table}
\end{figure*}

Similarly, we cast the ratchet controller as a \emph{transducer} that maps from
input sequences to distributions over output sequences.

\begin{Def}
\label{Def:Transducer}
A finite-state edge-labeled \emph{transducer} consists of:
\begin{enumerate}
\setlength{\topsep}{0mm}
\setlength{\itemsep}{0mm}
\item A finite set of states
	$\RatchetStateAlphabet = \{\RatchetStateVar_1, ... , \RatchetStateVar_{N'} \}$,
\item A finite input alphabet $\MeasAlphabet$ of $k$ symbols
	$\msym \in \MeasAlphabet$,
\item A finite output alphabet $\MeasAlphabet'$ of $k'$ symbols
	$\msym' \in \MeasAlphabet'$,
and,
\item A set of $N'$ by $N'$ input-output symbol-labeled transition matrices	
	$T^{(\msym, \msym')}$, $\msym, \msym' \in \MeasAlphabet \times \MeasAlphabet'$:
\begin{align*}
    T^{(\msym, \msym')}_{ij}
	  = \Pr(\RatchetStateSym_j,\msym'|\RatchetStateSym_i, \msym)
	~.
\end{align*}
\end{enumerate}
\end{Def}

The transducer formulation allows us to calculate the output HMM in terms of the
ratchet and the input machine. The exact method is given in
\cref{app:CompositionMJ}. As with the input machine, a ratchet is memoryless
when it possesses only one internal state, and memoryful otherwise. A key
feature of a ratchet we focus on here is its ability to alter temporal
correlations by altering the structure of an input process. If memoryless (IID)
input is fed to a memoryful ratchet, generally the output will be memoryful,
since this guarantees the state space dimension of the output
$|\RatchetStateAlphabet'| = |\InputStateSet \times \RatchetStateAlphabet| > 1$.
Figure \ref{fig:MJ_comp_table} graphically illustrates the composition of
various input process HMMs with the ratchet transducer we analyze in detail
shortly.

\subsection{Informatics}
\label{sec:Information Engine Informatics}

Following Landauer, extensions of the Second Law of Thermodynamics were
proposed to bound the thermodynamic costs of information processing by an
information engine. Reference \cite{Mand012a} employed a bound that compares
the \emph{Shannon entropy} of single input and single output symbols. Recall
that the Shannon entropy $H_1$ for a single random variable $X$ realizing
values $x \in \mathcal{X}$ is:
\begin{align}
  H_1 [X] = - \sum_{x \in \mathcal{X}} \Pr(X = x) \log_2 \Pr(X = x)
    ~.
  \label{eq:H_1}
\end{align}
This Shannon entropy quantifies the randomness of the single random variable $X$
averaged over time. That is, $H_1[\TapeVar]$ answers the question of
how uncertain it is that any particular $\TapeSym_N$ will be $0, 1,
\ldots,$ or $k-1$.

Comparing the single-symbol Shannon entropy in the input string to that in
the output string quantifies how the ratchet transforms randomness in individual
symbols. This difference captures one aspect of the ratchet's information
processing. And, it was proposed as an upper bound on the asymptotic work done
$\langle W \rangle$ \cite{Mand012a}:
\begin{align}
  \langle W \rangle & \stackrel{?}{\leq} \kB T \Delta H_1
    ~,
  \label{eq:SSBound}
\end{align}
where $H_1 = H_1[\TapeVar]$ is the entropy averaged over the input tape, $H_1'
= H_1[\TapeVar']$ is that averaged over the output, and $\Delta H_1 = H_1' -
H_1$ is the change in the single-symbol statistics produced by the ratchet's
operation.

Note, however, that while the $H_1$s track the average information in any single
instance of $X_t$ or $X'_t$, they do not account for temporal correlations
within input sequences or within output sequences. This is key, as information
ratchets change more than the statistical bias in an individual symbol, they
alter temporal correlations in symbol strings. These altered correlations are
related to the fact that the ratchet induces structural change in its
input. Recognizing this is central to bounding the thermodynamic costs of the
ratchet's interaction with the input process.

To properly address how correlations affect costs, we calculate a process'
intrinsic randomness when all temporal correlations are taken into account, as
measured by the \emph{entropy rate} \cite{Cove91a}:
\begin{align}
  \hmu = \lim\limits_{\ell \rightarrow \infty} \frac{H(\ell)}{\ell}
  ~,
\label{eq:hmu}
\end{align}
where $H(\ell) = \H[\Prob(\MS{0}{\ell-1})]$ is the Shannon entropy for
length-$\ell$ symbol blocks.

Replacing the input and output Shannon entropies in Eq. (\ref{eq:SSBound}) with
their respective entropy rates gives the \emph{Information Processing Second
Law} (IPSL) \cite{Boyd15a}:
\begin{align}
\langle W \rangle & \leq \kB T \ln 2 \left( \hmu' - \hmu \right) \nonumber \\
  & = \kB T \ln 2 ~ \Delta \hmu
  ~.
\label{eq:IPSL}
\end{align}
The IPSL correctly expresses the upper bound on work, taking into account the
presence of temporal correlations in input and output processes.

The importance of \cref{eq:IPSL} cannot be overstated---any memoryful ratchet
induces temporal correlations in its output, even for IID input. Using
\cref{eq:SSBound} in the IID case typically overestimates the upper limit on
available work. Additionally, temporal correlations in the input are known to be
a thermodynamic resource \cite{Boyd16d}. In fact, suitably designed ratchets can
leverage such correlation to do useful work. Thus, inappropriately applying
\cref{eq:SSBound} in these cases often results in claims that violate the
Second Law. In short, \cref{eq:IPSL} generalizes Landauer's Principle to the
case of correlated environments and finite-state memoryful ratchets that
generate correlated outputs.

Unfortunately, due to the difficulty of accurately calculating the entropy rate
for most processes, previous treatments of information ratchets were restricted
to use either \cref{eq:SSBound} or finite-length approximations to
\cref{eq:IPSL}. Here, we make use of a novel solution that removes this
restriction and gives accurate calculations of entropy rates for processes
generated by general HMMs.

\section{Entropy Rate of HMMs}
\label{sec:CalculationofEntropyRate}

Properly determining the entropy rate of processes generated by HMMs is a
longstanding challenge, one known since the 1950s \cite{Blac57b}. Its recent
resolution required introducing new concepts from ergodic theory and dynamical
systems \cite{Jurg19a}. We now turn to briefly discuss these and the new
analysis tools that follow from them. (The Supplementary Materials give a more
detailed exegesis.)

\subsection{$\epsilon$-Machines}
\label{sec:EpsilonMachines}

First, though, we need to more carefully consider the hidden Markov models that
we use to represent stochastic processes. We briefly recall two important
HMM classes.

\begin{Def}
A \emph{unifilar HMM} (uHMM) is an HMM such that for each state $\causalstate_k
\in \CausalStateSet$ and each symbol $\msym \in \MeasAlphabet$ there is
at most one outgoing edge from state $\causalstate_k$ labeled with symbol
$\msym$.
\label{def:UHMM}
\end{Def}

This seemingly-minor structural property means that the states are
\emph{predictive}: current state and symbol exactly predict the next state.
This has important consequences for calculating the statistical and
informational properties of the process that an HMM generates. If an HMM is
unifilar, we may directly calculate the entropy of its generated process via
the closed-form expression:
\begin{align}
\hmu = - \sum_{\cs \in \CausalStateSet} \Pr(\cs) \sum_{\cs' \in \CausalStateSet, \msym \in \MeasAlphabet} T_{\cs \cs^\prime}^{(\msym)} \log_2 T_{\cs \cs^\prime}^{(\msym)}
  ~.
\label{eq:ShannonEntropyRateHMM}
\end{align}
In contrast, if an HMM is nonunifilar, its states are not predictive and there
is no closed form for the generated process' entropy rate.

\begin{Def}
An \emph{\eM} is a uHMM with \emph{probabilistically distinct states}: For each pair of distinct states $\causalstate_k, \causalstate_j \in \CausalStateSet$ there exists some finite word $w = \ms{0}{\ell-1}$ such that:
\begin{align*}
\Prob(\MS{0}{\ell} = w|\CausalState_0 = \causalstate_k)
  \not= \Prob(\MS{0}{\ell} = w|\CausalState_0 = \causalstate_j)~.
\end{align*}
\label{def:eM}
\end{Def}

As a consequence, a process' $\eM$ is its optimally-predictive model
\cite{Crut12a}. Moreover, a process' $\eM$ is minimal and unique. This means
that we can quantify the amount of structural memory a process effectively uses
by counting the number of states in its $\eM$ or by calculating its stored
information. The latter is the \emph{statistical complexity} $\Cmu$, which
is the Shannon entropy of the asymptotic probability distribution over states:
\begin{align}
  \Cmu & = \H[\Pr(\CausalState)] \nonumber \\
  & = - \sum_{\causalstate \in \CausalState} \Prob(\causalstate) \log_2 \Prob(\causalstate) ~.
\label{eq:Cmu}
\end{align}
So, knowing a process' $\eM$ is powerful, as it provides closed-form
expressions for both a process' intrinsic randomness and its structural memory
\cite{Crut13a}, two important thermodynamic resources.

That said, even if a ratchet's input is generated by a finite $\eM$, the output
process will not be. In general, the output process generator (Fig.
\ref{fig:MJ_comp_table}, right column) will be a nonunifilar HMM. This
precludes a direct calculation of the entropy rate of a ratchet's output
process. And so, when determining thermodynamic function, it appears that a key
constituent ($\hmu'$) is inaccessible. Note, too, that nonunifilarity precludes
determining the output process' memory $\Cmu^{'}$ and, failing that, one cannot
accurately analyze the changes in structure effected by a ratchet.

\begin{figure*}
  \centering
  \includegraphics[width=0.9\textwidth]{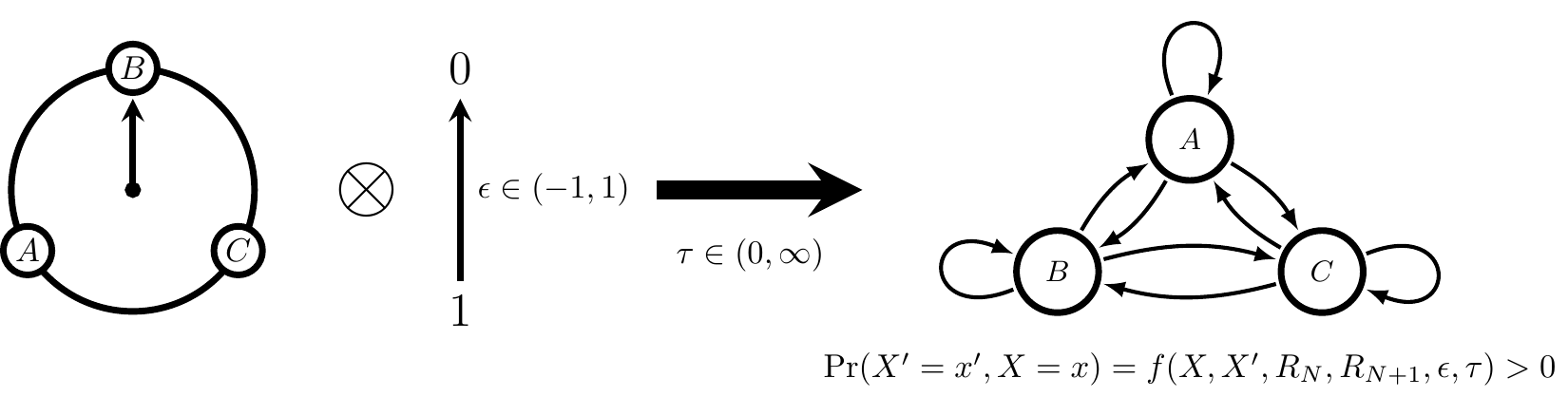}
  \caption{Ratchet schematic adapted from the original Mandal-Jarzynski
    construction, showing how the dial-and-symbol system is transformed into a
    three state transducer upon selection of a specific
    $\WeightParameter$---determining the energetics of flipping a bit---and
    $\tau$---determining the interaction interval. For almost every value of
    $\WeightParameter$ and $\tau$ every state-to-state transition has positive
    probability for every input-output symbol combination.}
  \label{fig:MJ_composition}
  \end{figure*}

\subsection{Mixed-State Presentation}
\label{sec:MixedStatePresentation}

Despite there being no closed-form expression for the entropy rate of the
process generated by a finite nonunifilar HMM, there is a way to
\emph{unifilarize} HMMs, introduced by Blackwell \cite{Blac57b}, using
\emph{mixed states}. A mixed state is the answer to the question: ``given that
one knows the HMM structure (states and transitions) and has observed a
particular sequence, what is the best guess of the internal state
probabilities?'' More formally, an $N$-state HMM's \emph{mixed states} are
conditional probability distributions $\eta(\meassymbol_{-\ell:0}) =
\Pr(\AlternateState_0 | \MeasSymbol_{-\ell:0} = \meassymbol_{-\ell:0}$) over the
HMM's internal states $\AlternateState$, given all sequences
$\meassymbol_{-\ell:0} \in \Abet^\ell$, $\ell = 0, 1, 2, \ldots$.

The collection of mixed states over all of a process' allowed sequences, i.e.,
$\ell \to \infty$, induces a (\emph{Blackwell}) measure $\mu$ on the state
distribution $\Pr(\AlternateState)$ $(N-1)$-dimensional simplex $\MxSSet$. The
mixed states together with the mixed-state transition dynamic (see SM
\cref{app:MixedStates}) give an HMM's \emph{mixed-state presentation} (MSP).

The MSP is unifilar by construction. However, in the typical case, this
improvement comes at a heavy cost---the set of mixed states is uncountably
infinite. This renders the complexity-measure expressions
\cref{eq:ShannonEntropyRateHMM} and \cref{eq:Cmu} unusable. Blackwell provided
a formal replacement for \cref{eq:ShannonEntropyRateHMM}'s entropy-rate
expression \cite{Blac57b}. This is an integral expression for the entropy rate
over the invariant Blackwell measure $\mu(\mxst)$ in the mixed-state simplex
$\MxSSet$:
\begin{align}
  h^B_{\mu} = - \int_{\MxSSet} d \mu(\mxst) \sum_{x \in \MeasAlphabet} \Pr( x | \mxst) \log_2 \Pr(x | \mxst) ~.
\label{eq:BlackwellHmuIntegral}
\end{align}

Recently, Ref. \cite{Jurg19a} introduced a constructive approach to evaluate
this integral by establishing contractivity of the simplex maps---the
substochastic transition matrices of Def. \ref{Def:HMM}---by showing that the
mixed-state process is ergodic. Given this, rather than integrate over the
measure $\mu(\mxst)$ as required by Blackwell's Eq.
(\ref{eq:BlackwellHmuIntegral}), we can then time-average over the series
$\mxst_0, \mxst_1, \ldots, \mxst_t$ of iterated mixed states to obtain the
entropy rate:
\begin{align}
  \widehat{\hmu^B} = - \lim_{\ell \to \infty}
  \frac{1}{\ell} \sum_{t = 0}^\ell
  \sum_{\msym \in \MeasAlphabet}
  \Pr(\msym |\mxst_\ell) \log_2 \Pr(\msym |\mxst_\ell)
  ~,
\label{eq:BlackwellHmuSum}
\end{align}
where $\Pr(\msym |\mxst_\ell) = \mxst(\ms{0}{\ell}) \cdot T^{(\msym)} \cdot
\One$, $\ms{0}{\ell}$ is the first $\ell$ symbols of an arbitrarily long
sequence $\ms{0}{\infty}$ generated by the mixed-state process, and $\One$ is a
column-vector of all $1$s. (See SM \cref{app:MixedStates}.) Applying
\cref{eq:BlackwellHmuSum} in the present setting means we can now calculate the
entropy rate of output processes for arbitrary ratchets and arbitrary inputs.

Characterizing a nonunifilar HMM's structure is slightly more delicate. Due to
the generic uncountability of predictive states for nonunfilar HMMs, $\Cmu$ of
the set of mixed states diverges. To characterize the divergent memory resource
cost of predicting processes with uncountably infinite mixed states, we track
the rate of divergence of $\Cmu$---the \emph{statistical complexity dimension}
$\CmuDim$ of the Blackwell measure $\mu$ on $\MxSSet$ \cite{Marz17a}:
\begin{align}
  \CmuDim = \lim_{\CoarseGrain \to 0}
  - \frac{\H_\CoarseGrain [\AlternateState]}{\log_2 \CoarseGrain}
  ~,
\label{eq:CmuDim}
\end{align}
where $\H_\CoarseGrain [Q]$ is the Shannon entropy of a continuous-valued
random variable $Q$, coarse-grained at size $\CoarseGrain$, and
$\AlternateState$ is the random variable associated with the mixed states $\eta
\in \AlternateStateSet$. SM \cref{app:MixedStates} develops an upper bound on
this that can be accurately determined from the measured process' entropy rate
$\widehat{\hmu^B}$ (\cref{eq:BlackwellHmuSum}) and the mixed-state process'
Lyapunov characteristic exponent spectrum $\Lambda$. As discussed in SM
\ref{app:CmuDim}, this upper bound is a close approximation to $\CmuDim$ for
broad classes of HMMs, but may be a strict inequality for others.

In this way, the functional thermodynamics of finite-state Maxwellian ratchets
can be accurately determined and systematically explored.

\section{Mandal-Jarzynski Information Ratchet}
\label{sec:MJRatchet}

\begin{figure*}
  \centering
  \includegraphics[width=\textwidth]{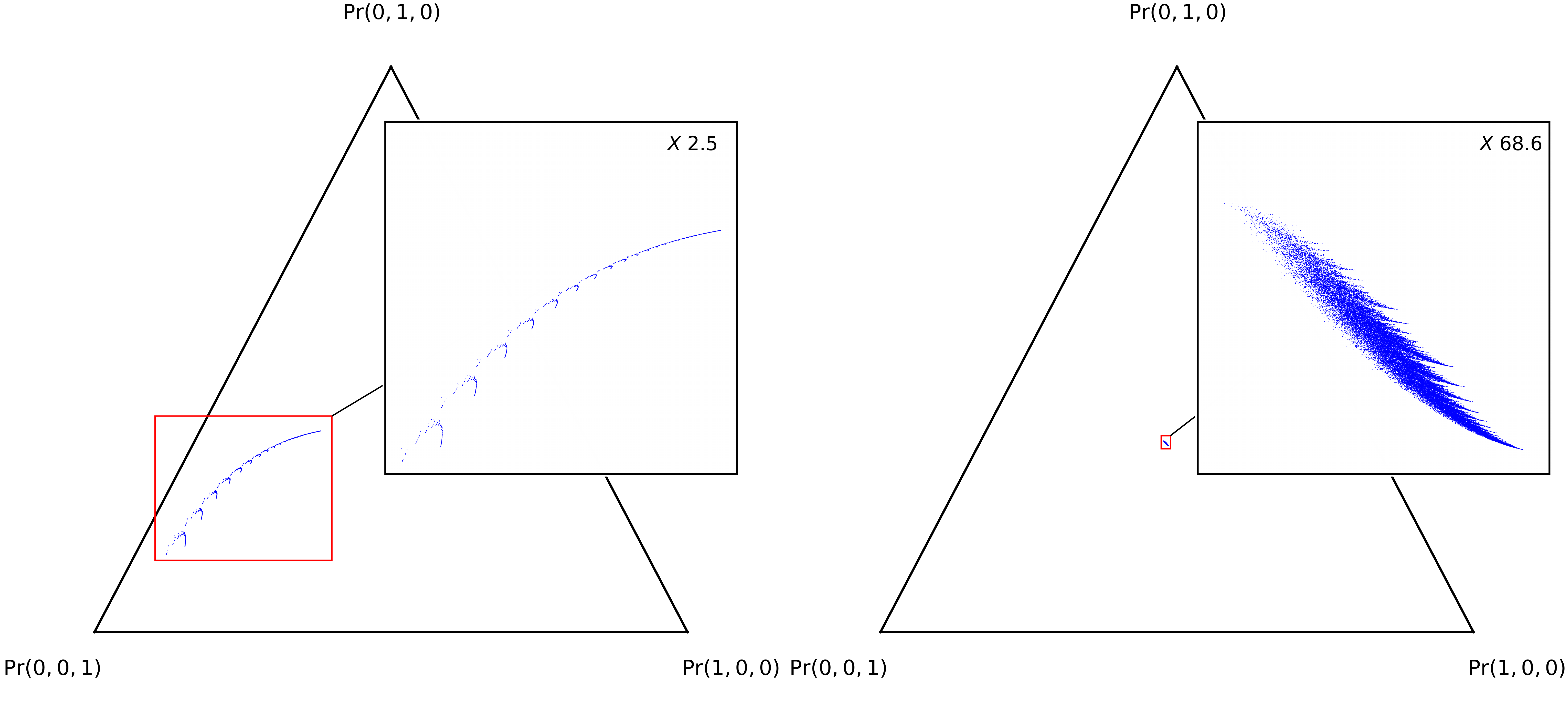}
  \caption{Mixed states $\eta \in \MxSSet$ of the output process generated by the
    ratchet driven with memoryless input (\cref{fig:MJ_comp_table}(top row))
    plotted on the $2$-simplex. Corner labels give the mixed-state
    probability distributions $\eta = \left(\Pr(A\otimes D), \Pr(B\otimes D),
    \Pr(C\otimes D) \right)$. Mixed states at the simplex corners correspond to
    the HMM being in exactly one of its states, while mixed states in the
    simplex interior are mixtures of the possible HMM states, with $\eta =
    \left( \frac{1}{3}, \frac{1}{3}, \frac{1}{3}\right)$ lying at the center.
    (Left) Ratchet parameters $\delta = -0.98$, $\WeightParameter = 0.01$, and
    $\tau = 0.1$. (Right) Ratchet parameters $\delta = 0.4$, $\WeightParameter
    = 0.5$, and $\tau = 0.1$. Insets: Detail of the mixed-state sets, magnified
    by amount indicated in upper right corner.
    }
  \label{fig:MJ_simplex}
  \end{figure*}

To demonstrate the descriptive power of these dynamical-thermodynamic results
on ratchet entropy, dimension, mixed states, and function, we apply them to a
well-known example of an information engine---the Mandal-Jarzynski ratchet
\cite{Mand012a}; hereafter, \emph{the ratchet}. Although initially introduced
without reference to HMMs and transducers, following Ref. \cite{Boyd15a} we
translate the original ratchet model into the HMM-transducer formalism outlined
in \cref{sec:InfoEngines}. In these terms, the ratchet is a three-state, fully
connected transducer, designed such that only transitions that flip an incoming
symbol are energetically consequential. As shown in \cref{fig:MJ_composition},
the ratchet's transition probabilities are parametrized by $\tau \in \left[ 0,
\infty \right)$---duration of the ratchet-symbol interaction---and
$\WeightParameter \in \left( -1, 1\right)$---the \emph{weight parameter}. For a
given $\tau$ and $\WeightParameter$, the Mandal-Jarzynski model may be written
down as the three-state transducer shown in the center column
\cref{fig:MJ_comp_table}. See SM \cref{app:MandalJarzynski} for how to
calculate the transducer, which is based on a rate-transition matrix, and SM
\cref{app:CompositionMJ} for the input-transducer composition method.

Any interaction interval in which the input symbol is unchanged is
energetically neutral. Therefore, we measure the average work done by the
ratchet by the difference in the probability of reading a $1$ on
the input tape cell versus writing a $1$ to the output tape cell:
\begin{align*}
  \langle W \rangle = \kB T
  ~ w \left( \Pr(\TapeVar' = 1) - \Pr(\TapeVar = 1)\right)
  ~,
\end{align*}
where $w(\WeightParameter) = \log((1+\WeightParameter)/(1-\WeightParameter))$.
When $\WeightParameter = 0$, flips $0 \to 1$ and $1 \to 0$ are both
energetically neutral; when $\WeightParameter \to \pm 1$, symbol flips in one
direction are energetically favored over the other. Note that this computation
finds the same asymptotic work production as \cref{eq:WorkProduction};
recalled here as an aid to intuition. 

Reference \cite{Mand012a}'s initial analysis considered only uncorrelated
inputs. That is, their input machine was a single-state HMM---a biased coin,
with bias $\delta = \Pr(0)-\Pr(1)$. To identify their ratchet's thermodynamic
functionality, the work bound was approximated via
\cref{eq:SSBound}---that is, assuming tape symbols were statistically
independent. However, the ratchet is memoryful (due to its three internal
states) and, therefore, in general induces correlations in its output, even for
uncorrelated inputs. Since the single-symbol entropy only upper-bounds the true
Shannon entropy rate---$\hmu \leq H_1$---\cref{eq:SSBound} is suspect when used
to identify actual thermodynamic functioning. Using new results here, the
following shows that, while approximately correct for uncorrelated input, the
single-symbol entropy bound is violated for correlated input. Its incorrect use
mischaracterizes thermodynamic functioning and can lead to violations of the
Second Law.

In addition, our new methods give insight into how the ratchet processes
structural information. Due to its inherent nonunifilarity, even when driven by
a finite-state $\eM$, the ratchet produces nonunifilar output machines that
generate processes with an uncountably-infinite set of mixed states, as
\cref{fig:MJ_simplex} shows. Moreover, the figure also demonstrates that as
ratchet parameters vary the mixed-state sets have strikingly different
structure.

Previous interpretations of ratchet thermodynamic functioning were limited to
considering only transformations of randomness; i.e., for given ratchet
parameters and input, what is the sign and magnitude of $k_B T \ln 2 ~\Delta
\hmu$ and how does this affect $\langle W \rangle$? Such questions ignore the
key second dimension of information processing illustrated so vividly by
\cref{fig:MJ_simplex}. That is, given the same ratchet, parameters, and input,
what is the sign and magnitude of $\Delta \Cmu$ and $\Delta d_\mu$? Does the
ratchet construct new patterns in its output ($\Delta \Cmu > 0$ or $\Delta
d_\mu > 0$) or deconstruct patterns passed to it from the input ($\Delta \Cmu <
0$ or $\Delta d_\mu < 0$)? How do these then affect $\langle W \rangle$?
Answering structural questions requires a more thorough taxonomy of
thermodynamic functionality than the original engine/dud/eraser categories.

\section{Randomizing and Derandomizing Behaviors}
\label{sec:RandomizingAndDerandomizingBehaviors}

The ratchet's previously-identified thermodynamic functions \emph{engine},
\emph{eraser}, and \emph{dud} were identified by comparing the sign and
magnitude of $k_B T \ln 2 ~\Delta \hmu$ to the asymptotic work production.  As
such, there are three physically possible orderings:
\begin{itemize}
      \setlength{\topsep}{-2pt}
      \setlength{\itemsep}{-2pt}
      \setlength{\parsep}{-2pt}
      \setlength{\labelwidth}{5pt}
      \setlength{\itemindent}{0pt}
\item Engine: $ 0 < \langle W \rangle \leq k_B T \ln 2 ~\Delta \hmu$;
\item Eraser: $ \langle W \rangle \leq k_B T \ln 2 ~\Delta \hmu < 0$; and
\item Dud: $ \langle W \rangle \leq 0 \leq k_B T \ln 2 ~\Delta \hmu$.
\end{itemize}
A ratchet randomizing inputs ($\Delta \hmu > 0$) can operate as an engine, if
it is leveraging the change in entropy rate to do useful work. It may also act
as a dud, if the randomization produces no useful work or, worse, if the
ratchet is using work. A ratchet derandomizing inputs ($\Delta \hmu < 0$) is
termed an ``eraser'' and can only derandomize up to $\langle W \rangle / k_B T
\ln 2$ bits using $\langle W \rangle$ joules of work. The ordering $k_B T \ln 2
~\Delta \hmu < \langle W \rangle < 0$ would imply that the ratchet is
derandomizing beyond the physical limitations of Landauer's principle. 

As noted already, Ref. \cite{Mand012a} originally identified these
functionalities using the entropy-change approximation $\Delta H_1$ rather than
the exact change $\Delta \hmu$ introduced by Ref. \cite{Boyd15a}. As previously
shown, driving a memoryful ratchet with a memoryful input violates
\cref{eq:SSBound} \cite{Boyd16d}. In all other cases, \cref{eq:SSBound} is
valid, but may mischaracterize the functional thermodynamic regimes. A natural
question, therefore, is how much difference does using the correct entropy rate
make in identifying function? To see this, we now compare $\Delta \hmu$ and
$\Delta H_1$.

There are three possibilities. First, $\Delta H(1) = \Delta \hmu$. In this case,
a ratchet does not change the presence of temporal correlations. This occurs
when a memoryless ratchet is driven by memoryless input.

Second, $\Delta H(1) > \Delta \hmu$. Here, a ratchet reduces the presence of
temporal correlations, which occurs when a memoryless ratchet has been driven
by memoryful input. In this regime, the difference in single-symbol entropy is
a tighter bound on the correlation change than the difference in entropy rate.
Critical to this case, though, recall that our goal is not a tight bound, but
rather an accurate measurement of the gap between information processing and
asymptotic work. The upshot is that using \cref{eq:SSBound} in this case may
mischaracterize thermodynamic functionality.

Finally, $\Delta H(1) < \Delta \hmu$, which occurs when a memoryful ratchet is
driven by memoryless input. In this case, the ratchet increases temporal
correlations in the output, so that the difference in entropy rates is a tighter
bound on the asymptotic work production. This is the scenario in the first
treatment of the Mandel-Jarzynski ratchet \cite{Mand012a}. Note that when a
memoryful ratchet is driven with memoryful input, the most generic case, all
orderings of $\Delta \hmu$ and $\Delta H(1)$ are possible.

Let's now turn to consider in detail how the ratchet operates in three distinct
environments: Memoryless, periodic, and memoryful inputs. This gives more
direct insight into the ratchet's transformational capabilities.

\subsection{Memoryless Input}
\label{sec:MemorylessInput}

When the ratchet is driven with a memoryless input, as in the original analysis,
\cref{eq:SSBound} is valid, but IPSL always offers
a tighter or equal bound on work production than the single-symbol entropy
approximation. This holds since the input is memoryless, while the
three-state output machine is memoryful and nonunifilar for
almost every parameter setting. As such, one cannot calculate the entropy rate
$\hmu'$ in closed form. However, the new techniques above can
determine the mixed-state presentations of the output HMMs and this gives
accurate numerical calculation of both the single-symbol and the IPSL
work bounds. 

\begin{figure*}
  \centering
  \includegraphics[width=\textwidth]{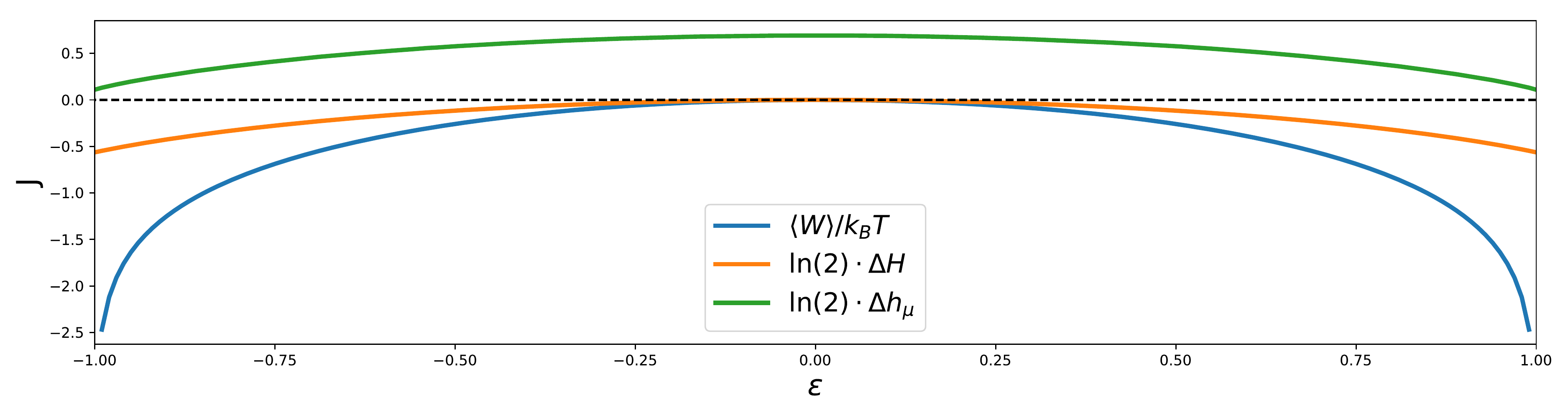}
  \caption{Asymptotic work production $\langle W \rangle$, single-symbol $\Delta
    H_1$ bound, and Kolmogorov-Sinai-Shannon $\Delta \hmu$ bound when ratchet is
    driven by period-2 memoryful input. Since the input has no parameters, the
    parameter sweeps only over $\WeightParameter$ with $\tau = 10$.}
  \label{fig:period_2_work}
  \end{figure*}

This all being said, for most parameter values of the Mandal-Jarzynski ratchet,
in practice we find that $\Delta \hmu \approx \Delta H_1$. In other words, when
driven with a memoryless input, the ratchet's functional thermodynamic regions
are not significantly changed when identified via the single-symbol entropy---a
minor quantitative difference without a functional distinction. (See
\cref{fig:MJ_tau_1} for a comparison of the functional thermodynamic regions
found by each bound.) Exploring output-machine MSPs shows this arises from the
ratchet's transition topology. As shown in the middle column of
\cref{fig:MJ_comp_table}, the ratchet's transducer is fully connected, and all
transitions to any other state on any combination of symbols are possible.
Therefore, it is impossible to be certain about which state the ratchet is in;
or, indeed, to even be sure which states the ratchet is \emph{not} in.
Graphically, this is represented by the fact that the output-machine mixed
states $\mxst$ always lie deep in the simplex $\MxSSet$'s interior, as
illustrated in Fig. \ref{fig:MJ_simplex} (Right).

Mixed states lying at $\MxSSet$'s center correspond to an equal belief in each
of the output HMM's three states: $A \otimes D$, $B \otimes D$, and $C \otimes
D$. While those on $\MxSSet$'s border indicate certainty of \emph{not} being in
at least one state. Since the probability distribution over the next symbol is a
continuous function over the mixed states (see \cref{app:MixedStates}), the
diameter of the mixed-state set is a rough measure of the presence of temporal
correlations in the ratchet's behavior. To explicitly illustrate this, the mixed
states for two example output processes generated by the ratchet are shown in
\cref{fig:MJ_simplex}. On the left, the mixed states are spread out, indicating
that at the selected parameters, the ratchet induces stronger temporal
correlations than in the next example (right). There, all mixed states lie very
close together and very near the simplex center. The mixed-state set has very
small diameter. For most parameter values, one finds that the mixed states of
the memoryless-driven ratchet's output process cluster closely in the middle of
the simplex. (See \cref{fig:att_grid} for a broader survey of ratchet MSPs for
memoryless input.) So, by giving insight into the mixed states of the output
process, our new techniques rather directly explain why $\Delta \hmu \approx
\Delta H_1$.

\subsection{Periodic Input}
\label{sec:PeriodicInput}

Now, consider driving the ratchet with a periodic input. The \emph{Period-$2$
Process}, shown in the middle row of \cref{fig:MJ_comp_table}, is memoryful,
with two internal states. So, it now is possible that \cref{eq:SSBound} is
violated. Since $H_1 = 1$ and $\hmu = 0$, the presence of temporal correlations
in the input is maximized. Noting this, and the near-memoryless behavior of the
ratchet as discussed in \cref{sec:MemorylessInput}, we can see that for almost
all parameters, the ratchet decreases the presence of temporal correlations in
transforming the input process to the output. The periodically-driven ratchet
output HMMs have six states and are nonunifilar for nearly all parameter values;
see \cref{fig:MJ_comp_table} (middle, last column). And so, we must calculate
the these machine's mixed-state presentations to estimate $\hmu'$. Comparing
\cref{eq:SSBound} and \cref{eq:IPSL} in \cref{fig:period_2_work} to the
asymptotic work production shows that \cref{eq:SSBound} is not violated. As
predicted above, it is a tighter bound on $\langle W \rangle$ than
\cref{eq:IPSL}.

Although it may seem desirable to use the tighter bound, the single-symbol and
entropy rate bounds identify the ratchet's thermodynamic functioning
differently: Since $\langle W \rangle \leq k_B T \ln 2 ~ \Delta H_1 \leq 0$ for
all values of $\WeightParameter$, the single-symbol entropy bound classifies the
ratchet as an eraser, dissipating work to reduce the randomness in the input.
However, when considering temporal correlations, we see that the ratchet is in
plain fact a dud---$\Delta \hmu > 0 $. That is, the ratchet dissipates work
while \emph{increasing} the tape's intrinsic randomness. This marked
mischaracterization of thermodynamic function by the single-symbol entropy
highlights an important lesson: \emph{Bounding the asymptotic work production as
tightly as possible is not the same as correctly identifying the functional
thermodynamics}. As Ref. \cite{Boyd17a} recently showed, rather than merely a
bound, \cref{eq:IPSL} is meaningful only when comparing $k_B T \ln 2 ~\Delta
\hmu$ to $\langle W \rangle$. The difference in the two quantifies the amount of
work the ratchet can do, if it were an optimal, globally-integrated information
processor. This shows that even when it may appear to outperform \cref{eq:IPSL},
in general \cref{eq:SSBound} cannot serve as a reliable bound on asymptotic work
production. We return to this in our final example, where applying
\cref{eq:SSBound} implies a violation the Second Law.

\subsection{Memoryful Input}
\label{sec:MemoryfulInput}

\begin{figure*}
  \centering
  \includegraphics[width=\textwidth]{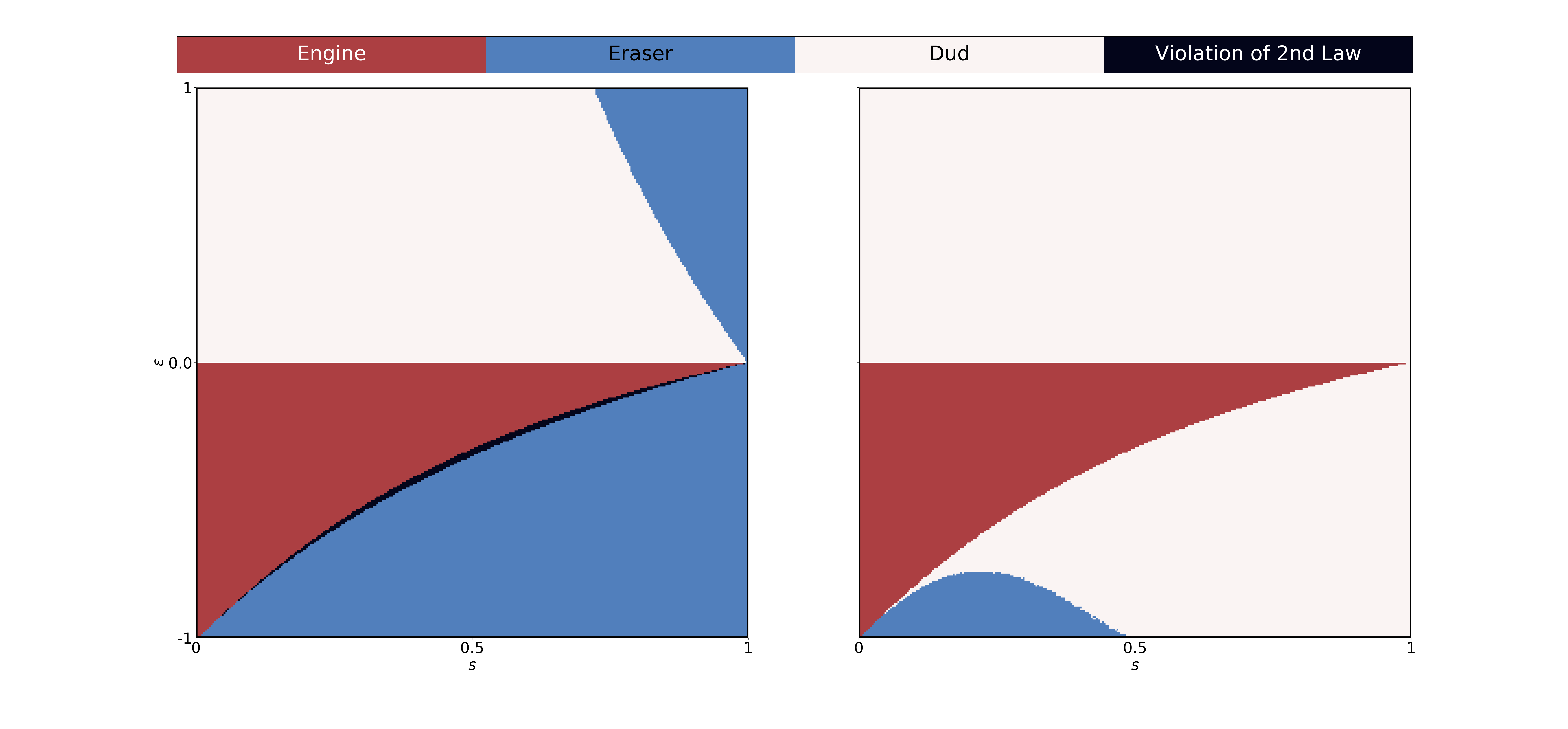}
  \caption{Functional thermodynamic regions of the ratchet driven with the
    Golden Mean Process as a function of parameters $\WeightParameter \in
    [-1,1]$ and $s \in [-1,1]$ with $\tau = 1$. (Left) Purported functionality
    identified via by single-symbol entropy bound \cref{eq:SSBound}. (Right)
    Correct functionality identified via the entropy-rate bound IPSL
    \cref{eq:IPSL}.
    }
  \label{fig:MJ_GM_tau_1}
  \end{figure*}

Finally, let's drive the ratchet with a mixed-complexity memoryful
process---partly regular, partly stochastic---the \emph{Golden Mean Process}. As
depicted in \cref{fig:MJ_comp_table} (bottom row), this two-state HMM generates
a family of processes parametrized by $s \in [0,1]$. When $s = 1$, the process
is period-$2$. Decreasing $s$ lets the process emit multiple $1$s in a row. This
increases in probability until at $s=0$, where the process emits only $1$s. The
driven ratchet's output HMMs have six states and are nonunifilar for nearly all
parameter values; see \cref{fig:MJ_comp_table}(bottom, last column). So, again,
we must calculate mixed-state presentations to get $\hmu'$ and identify
functionality.

In \cref{fig:MJ_GM_tau_1}, we apply both \cref{eq:SSBound} and \cref{eq:IPSL}
for the same set of ratchet parameters. For both, we find asymmetry in the
functional thermodynamic regions with respect to $\WeightParameter$, in
contrast to the highly symmetric regions found for memoryless input, shown in
\cref{fig:MJ_tau_1}. This is due to the asymmetry in input. In fact, it is not
possible for the Golden Mean Process to produce strings biased towards $0$.
Thermodynamically, for $\WeightParameter > 0$, the ratchet is not able to
extract work. When applying the single-symbol bound, as shown on the left in
\cref{fig:MJ_GM_tau_1}, the bound reports large regions of eraser behavior.
And, most importantly, between the engine and lower eraser region lies a region
where we see that \cref{eq:SSBound} implies
\begin{align}
  \langle W \rangle > k_B T \ln 2 ~\Delta H_1
  ~,
\end{align}
a violation of the Second Law!

Of course, when we apply \cref{eq:IPSL} in \cref{fig:MJ_GM_tau_1} (Right), the
violation region disappears, to be correctly identified as duds. Additionally,
the large region of eraser functionality in \cref{fig:MJ_GM_tau_1} (Left)
shrinks significantly in \cref{fig:MJ_GM_tau_1} (Right). \Cref{fig:MJ_GM_tau_1}
(Left)'s regions have been mischaracterized similar to the case discussed in
\cref{sec:PeriodicInput}. It is more subtle here, though, since $\hmu > 0$.
However, the fundamental problem is the same---by considering only the
single-symbol entropy, it appears that the ratchet performs work to make the
input less random, since $\Delta H_1 < 0$. In fact, the output is more
intrinsically random than the input, and the ratchet dissipates work uselessly.
In the violation region on the left, the ratchet is identified as not
dissipating sufficient work to reduce the randomness as much as $\Delta H_1$
implies it must be. This leads to the Second Law violation. This contradiction
is resolved when we take into account that the input's intrinsic randomness was
actually much lower than its single-symbol entropy. And so, the apparent
decrease in randomness was in fact an increase.

It is already known that \cref{eq:SSBound} may be violated in cases of a
memoryful ratchet driven by memoryful input. However, the Mandal-Jarzynski
ratchet was not designed to find such a violation, as has been done previously
\cite{Boyd16c}. Rather, we find that driving a simple transition-rate based
ratchet with a mixed-complexity process creates regions of violation when
applying \cref{eq:SSBound}. Since such ratchets are common in application, and
any such ratchet will be highly stochastic by nature, for reasons further
discussed in \cref{app:MandalJarzynski}, we conclude that \cref{eq:SSBound} is
not suitable to be broadly applied. On the positive side, we see that the
dynamical-systems techniques introduced here apply broadly, giving consistent
and accurate characterizations stochastic-control information engines.

\section{Constructing and Deconstructing Patterns}
\label{sec:ConstructingAndDeconstructingPatterns}

\begin{figure*}
  \centering
  \includegraphics[width=\textwidth]{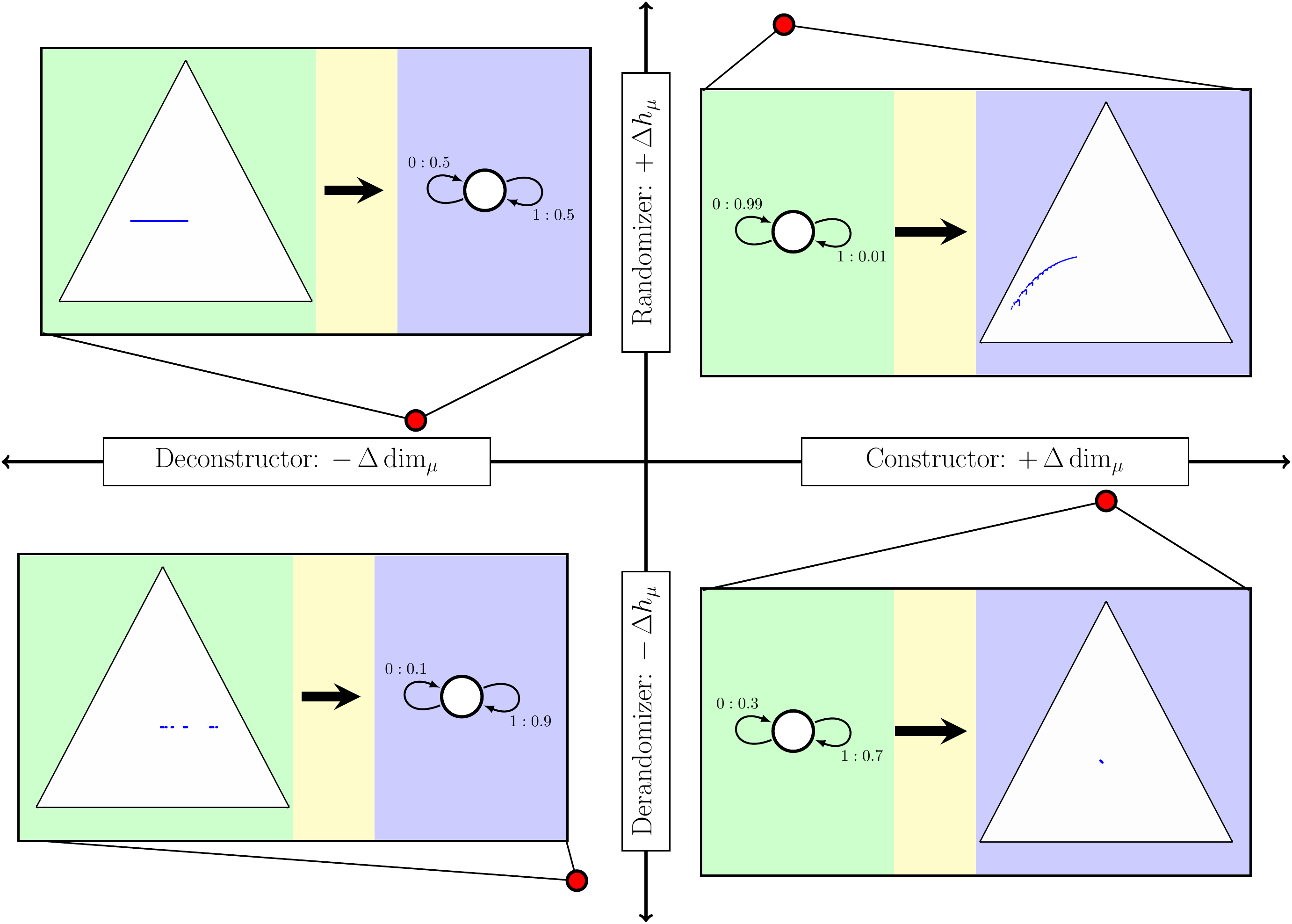}
\caption{After passing through the information engine, the input process,
	which has some initial $\hmu$ and $\CmuDim$, is transformed into an output
	process with a potentially different $\hmu'$ and $\CmuDim'$. By carefully
	selecting input, an information engine can be induced to act as a
	randomizer ($\Delta \hmu > 0$) or derandomizer ($\Delta \hmu < 0$) and a
	pattern constructor ($\Delta \CmuDim > 0$) or deconstructor ($\Delta \CmuDim
	< 0$). We show here that all four regions of the $\Delta \hmu$--$\Delta
	\CmuDim$ plane are accessible to the Mandal-Jarzynski ratchet by carefully
	selecting parameters and input.  The two insets on the left show the
	uncountable set of mixed states of an input process that the ratchet
	transduces to an IID output. The insets on the right show two
	uncountably-infinite-state output processes produced by running the
	Mandal-Jarzynski ratchet on a biased coin. The parameters, clockwise from
	top right: $\delta_{input} = -0.98, \WeightParameter = 0.01, \tau = 0.1$;
	$\delta_{input} = 0.3, \WeightParameter = 0.5, \tau = 0.1$;
	$\delta_{output} = 0.8, \WeightParameter = - 0.96, \tau = 0.75$;
	$\delta_{output} = 0.0, \WeightParameter = 0.9, \tau = 0.9$. 
    }
  \label{fig:HmuCmuPlot}
  \end{figure*}

Up to this point, we monitored how the ratchet changed the \emph{amount}
of intrinsic randomness present in a symbol sequence and leveraged this to do
useful work. When information ratchets are memoryful, they can alter not only
the statistical bias of a symbol sequence, but also the presence of temporal
correlations. This has thermodynamic consequences, as discussed above. Now, we
turn to consider by what \emph{mechanisms} an information ratchet changes the
presence of temporal correlations, which manifests in changes in sequence
structure and organization. 

By structure and organization, we refer the internal states of the HMM that
generates the input symbol sequence, the ratchet states and transitions, and
the output sequence. As depicted in \cref{fig:MJ_comp_table}, the input and the
ratchet each have their own set of internal states. Since the output machine is
the composition of the ratchet transducer and input HMM, its states are the
Cartesian product of the set of input states and set of output states. In
the simplest case, when a memoryless ratchet is driven by memoryless input,
there is only ever one state, and no temporal correlations are present at any
stage. The only possible action of the ratchet then is to change the
statistical bias of individual input symbols and transform this change in
Shannon entropy to a change in thermodynamic entropy.

When one or both of the input and ratchet are memoryful, the internal structure
of the output will be, in general, memoryful. That is, the ratchet has induced a
structural change in processing the input to generate the output. Consider two
basic structural-change operating modes \cite{Boyd17a}: \emph{pattern
construction}, where the output is more structured than the input, and
\emph{pattern deconstruction}, the output is less structured. As before, these
modalities are input-dependent---the same ratchet may exhibit either. Note that
structural change to the symbol sequence does not uniquely determine the
thermodynamic functionality associated with changes in randomness. It is
possible for an information engine to act as an engine, eraser, or dud while
constructing patterns. The same is true of deconstruction. Rather,
transformations of randomness and structure are orthogonal, and a ratchet's
information processing capabilities may lie anywhere in the $\Delta
\hmu$---$\Delta \CmuDim$ plane sketched in \cref{fig:HmuCmuPlot}.

\subsection{Pattern Construction}
\label{sec:PatternGeneration}

Ideal pattern construction occurs when a ratchet takes structureless input---an
IID process---to structured output. Therefore, when the ratchet is driven with a
biased coin input, it is operating as an ideal pattern constructor. As discussed
in \cref{sec:MemorylessInput}, driving the ratchet with memoryless input results
in an uncountably-infinite set of states in the output HMM for most parameter
values. The exception occurs along the line $\delta = \WeightParameter$
in parameter space, where the ratchet returns the input unchanged,
implying $\Delta \Cmu = 0$. At every other point in ratchet parameter
space $\Delta \Cmu = +\infty$ and the ratchet acts as a pattern constructor. As
can be seen from \cref{fig:MJ_tau_1}, this type of structural change can be
associated with any thermodynamic behavior. 

The resulting divergence of $\Cmu$ is a direct consequence of the nonunifilarity
induced by the ratchet. The structure generated by any ratchet driven by an IID
process is the set of mixed states of the ratchet, given knowledge of the
outputs. Due to the ratchet's topology, there is an uncountable infinity of such
mixed states. In this circumstance one uses the statistical complexity
dimension---$\CmuDim$ of \cref{eq:CmuDim}---of the set of output mixed states to
monitor the rate of the memory-resource divergence. $\Delta \CmuDim$
distinguishes between output machines with an uncountable infinity of states,
and so is able to compare the structural information processing of the ratchet
across parameter space.

\Cref{fig:HmuCmuPlot} places two examples of ideal pattern construction on the
right side of the $\Delta \hmu$--$\Delta \CmuDim$ plane with the associated
input and output machines, the latter plotted on the 2-simplex. Although it may
appear that the more entropic ratchet in the upper half of the plane constructs
a more ``complex'' pattern, this is not so. Refer back to \cref{fig:MJ_simplex}
and compare the dimension of the two sets of mixed states to see the opposite
is true. The ratchet operating in the $-\Delta \hmu$ half of the plane produces
a much denser set of states, resulting in a larger $\Delta \CmuDim$. In
addition to the structural transformation, the ratchet in the $+\Delta \hmu$
plane randomizes inputs as a dud, while the other derandomizes inputs as an
eraser. 

\subsection{Pattern Deconstruction}
\label{sec:PatternDeconstruction}

In a complementary fashion, the ratchet can deconstruct patterns. In ideal
pattern deconstruction, a ratchet transforms a memoryful input sequence, with
$\Cmu > 0$, to memoryless, IID output, with $\Cmu = 0$. When taking a
ratchet-focused view, as we do here, ideal pattern deconstruction is a more
involved task than ideal pattern construction, since we must carefully design
inputs that a ratchet will transform into a biased coin. Any correlations in
the input must be recognizable by the ratchet so that the ratchet can map them
to randomness. Similar to the previous discussion, we consider the induced
ratchet mixed states, but now we have knowledge of the inputs. The algorithm to
design the required input process, given knowledge of the ratchet, is discussed
in \cref{app:PatternDeconstruction}.

Critically, pattern deconstruction is not possible for all ratchet parameters
and desired output. That said, the Mandal-Jarzynski ratchet can perform as an
engine, eraser, or dud while deconstructing patterns, as can be seen in
\cref{fig:patt_ext}. As $\tau$ increases, the parameter-space region in which
the ratchet can extract patterns shrinks. At $\tau \to \infty$ pattern
extraction may only occur along the line $\delta = \WeightParameter$. In a
mirror of pattern construction, generating the input processes require
reference to the uncountably infinite set of mixed states of the ratchet.  In
general, this implies that an input process which maps to a memoryless output
process also has an uncountably-infinite set of states and $\Delta \Cmu \to -
\infty$. In other words, to properly ensure the output symbols are temporally
uncorrelated, the input process must remember its \emph{infinite past}. Once
again, the associated statistical complexity dimension $\CmuDim$---now of the
set of input mixed states---quantifies the rate of the memory-resource
divergence.

Two examples of ideal pattern deconstruction are placed on the left side of the $\Delta \hmu$--$\Delta \CmuDim$ plane in \cref{fig:HmuCmuPlot}, with the ratchet mixed states---on the 2-simplex---and the output machine. $\Delta \CmuDim$ is approximate, based on the dimension of the ratchet mixed states, which are conjectured to have the same dimension as the input mixed states.  

\subsection{Thermodynamic Taxonomy of Construction and Deconstruction}
\label{sec:ThermodynamicTaxonomy}

From its highly stochastic nature and from parameter sweeps like the one shown
in \cref{fig:att_grid}, we conclude that for almost all parameters, the
Mandal-Jarzynski ratchet is only able to construct patterns with infinite sets
of predictive features (mixed states). We conjecture that likewise, it is only
able to perfectly deconstruct infinite patterns. An interesting note is that
the input and output mixed-state sets and their dimensions are asymmetric. We
can visually see the asymmetry in \cref{fig:HmuCmuPlot}, which sketches the
$\Delta \hmu$--$\Delta \CmuDim$ plane and shows an example of the
Mandal-Jarzynski ratchet operating in all four quadrants. Infinite state output
constructed by the ratchet may span the simplex, but the mixed states of the
ratchet, while acting as a deconstructor, always lie along a line in the
simplex. This implies that while the ratchet may construct patterns up to
$\Delta \CmuDim = 2.0$, it is only able to deconstruct patterns up to $\Delta
\CmuDim = -1.0$. The difference in $\Delta \CmuDim$ in these two modalities
points to a difference in memory-resource divergence for pattern construction
versus pattern deconstruction. 

This asymmetry is not necessarily surprising. Recall the asymmetry in the
the ratchet's ability to randomize and derandomize behavior. The combined area
of dud and engine region in \cref{fig:MJ_GM_tau_1} consists of the ratchet's
randomizing regime, while the derandomizing regime is the comparatively small
eraser region. One interpretation of this asymmetry comes from the
thermodynamic limitations on the ordering of $\Delta \hmu$ and $\langle W
\rangle$: While an increase in $\Delta \hmu$ is thermodynamically unbounded,
$\Delta \hmu$ is constrained by the Second Law to only drop as low as the
minimum asymptotic work. This strongly suggests that there is a thermodynamic
taxonomy of structural transformation---one that parallels our existing
thermodynamic taxonomy of randomness transformation. We must leave finding such
a taxonomy and the analysis of more general ratchets with input-dependent
structural behavior to future work.

\section{Related Efforts}
\label{sec:Related}

We can now place the preceding methods and new results in the context of prior
efforts to identify the thermodynamic functioning of information engines. In
short, though, having revealed the challenge of exact entropy calculations and
the inherent divergence in structural complexity, the new methods appear to
call for a substantial re-evaluation of previous claims. We start noting a
definitional difference and then turn to more consequential comparisons.

The framework of information reservoirs discussed here differs from alternative
approaches to the thermodynamics of information processing, which include: (i)
active feedback control by external means, where the thermodynamic account of
the Demon's activities tracks the mutual information between measurement
outcomes and system state \cite{Kish2012, Sagawa12, Kundu12, Abreu12,
Vaikuntanathan11, Abreu11, Granger11, Horowitz11, Ponmurugan11, Toya10a,
Sagawa10, Cao04, Touchette00}; (ii) the multipartite framework where, for a set
of interacting, stochastic subsystems, the Second Law is expressed via their
intrinsic entropy production, correlations among them, and transfer entropy
\cite{Ito13, Hartich14, Horowitz14, Horowitz15}; and (iii) steady-state models
that invoke time-scale separation to identify a portion of the overall entropy
production as an information current \cite{Strasberg13, Esposito12}. A unified
approach to these perspectives was attempted in Refs. \cite{Horowitz14b,
Barato14, Horowitz13}.

These differences being called out, Maxwellian demon-like models designed to
explore plausible automated mechanisms that do useful work by decreasing the
physical entropy, at the expense of positive change in reservoir Shannon
information, have been broadly discussed elsewhere \cite{Strasberg13, Mand012a,
Mandal13, Barato13, Hoppenau14, Lu14, Um15}. However, these too neglect
correlations in the information-bearing components and, in particular, the
mechanisms by which those correlations develop over time. In effect, they
account for thermodynamic information-processing by replacing the Shannon
information of the components as a whole by the sum of the components'
individual Shannon informations. Since the latter is larger than the former
\cite{Cove06a}, using it can lead to either stricter or looser bounds than the
correct bound derived from differences in total configurational
entropies. Of more concern, though, bounds that ignore correlations can simply
be violated. Finally, and just as critically, the bounds refer to
configurational entropies, not the intrinsic dynamical entropy over system
trajectories---the Kolmogorov-Sinai entropy. A more realistic model was
suggested in Ref. \cite{Lu14a}. Issues aside, these designs have been extended
to enzymatic dynamics \cite{Cao15}, stochastic feedback control
\cite{Shiraishi15}, and quantum information processing \cite{Diana13, Chap15a}.

In comparison, our approach expands on that of Ref. \cite{Boyd15a} that
considers a Demon in which all correlations among the system components are
addressed and accounted for. As shown above, this has significant impact on the
analysis of Demon thermodynamic functionality. To properly account for
correlations, we developed a new suite of tools that allow quickly and
efficiently analyzing nonunifilar HMMs and related stochastic controllers, which removes the mathematical
intractability of analyzing correlations for arbitrary demons. We note that our
approach and results are consistent with the analyses that consider the entropy
of the system as a whole, therefore treating correlations in the system
implicitly, an approach epitomized by Ref. \cite{Parrondo15}. Since correlations
are not ignored, this approach is fully consistent with our treatment. This
being said, insofar as that work does not address specific partitioning of the
system, it does not offer an explicit accounting of the system's internal
correlations, as is done here. As previously discussed, one may derive
information ratchet-type results from that approach by considering an
explicit partitioning \cite{Boyd16d, Boyd17a, Riechers19}. While the results are
consistent, leaving the role of correlations implicit does not allow for
investigating how to best leverage them. It also does not give a way
to analyze internal computational structure. These remarks highlight the
importance of explicitly considering information engine-style partitioning. 

The dynamical-systems methods additionally allowed us to consider a Demon's
internal structure, which had only previously been investigated for unifilar
ratchets in Ref. \cite{Boyd17a}. From engineering and cybernetics to biology
and now physics, questions of structure and how an agent, here understood as
the ratchet, interacts with and leverages its environment---i.e., input---is a
topic of broad interest \cite{Ashby60, Ehrenberg80}. General principles for how
an agent's structure must match that of its environment will become essential
tools for understanding how to take thermodynamic advantage of correlations in
structured environments, whether the correlations are temporal or spatial.
Ashby's Law of Requisite Variety---a controller must have at least the same
variety as its input so that the whole system can adapt to and compensate that
variety and achieve homeostasis \cite{Ashby60}---was an early attempt at such a
general principle of regulation and control. For information engines, a
controller's variety should match that of its environment \cite{Boyd16d}.
Above, paralleling this, but somewhat surprisingly, we showed that for the
Mandal-Jarzynski ratchet to extract patterns from its environment, the input
must have an uncountably infinite set of memory states synchronized to the
ratchet's current mixed state. One cannot but wonder how such requirements
manifest physically in adaptive thermodynamic nanoscale devices and biological
agents.

\section{Conclusions}
\label{sec:Conclusions}

Thermodynamic computing has blossomed, of late, into a vibrant and growing
research domain, driven by applications and experiment
\cite{Maruyama09,Toya10a,Beru12a,Kish2012,Saga12a,Seif12a,Ito13,Strasberg13,Barato14,Shiraishi15,Cao15,Cont19a,Wims19a,Sair19c,Parrondo15}.
As such, it is vital that analytical tools accurately relate information
processing and thermodynamic functionality. While the original class of
Maxwellian information engines was flexible and well suited to specific
applications, accurate analysis and correct functional classifications were
previously hampered by the challenge of determining the entropy rate of
temporally-correlated sequences---sequences that are inevitably induced by
Maxwellian ratchets or are present in their possible environments. Previously
useful and seemingly reasonable approximations to the entropy rate are not up to
this task. As we demonstrated, they can fail miserably---even lead to incorrect
attributions of thermodynamic function and, worse, to violations of the Second
Law.

Here, we introduced new techniques from dynamical systems and ergodic
theory---dimension theory, iterated function systems, and random matrix
theory---that overcome these hurdles and, in the process, constructively solve
Blackwell's long-standing question of the entropy rate of processes generated
by hidden Markov models.  They allow us to accurately determine the
thermodynamic functioning of Maxwellian information engines with arbitrary
ratchet design, over all possible inputs. In this way, the results
significantly expand the set of analyzable engines. In short, this changes the
perspective of the current research program from studying highly constrained
toy examples to broadly surveying engine designs. This is a boon to both
theory, experiment, and engineering.

Furthermore, these tools allowed us to look under the hood, so to speak, and
examine more than quantitative changes in the intrinsic randomness of
processes, but also to show how ratchets impact structure and correlation. Most
strikingly, we showed that, in general, stochastic ratchets generate outputs
that require uncountably infinite sets of predictive features to optimally
function, even when driven by trivial (temporally uncorrelated) input.

\section*{Acknowledgments}
\label{sec:acknowledgments}

The authors thank Alec Boyd, Sam Loomis, Mikhael Semaan, and Ariadna Venegas-Li
for helpful discussions and the Telluride Science Research Center for
hospitality during visits and the participants of the Information Engines
Workshops there. JPC acknowledges the kind hospitality of the Santa Fe
Institute, Institute for Advanced Study at the University of Amsterdam, and
California Institute of Technology for their hospitality during visits. This
material is based upon work supported by, or in part by, FQXi Grant number
FQXi-RFP-IPW-1902, and U.S. Army Research Laboratory and the U.S. Army Research
Office under contract W911NF-13-1-0390 and grant W911NF-18-1-0028.

\appendix

\onecolumngrid
\clearpage
\begin{center}
{\huge Supplementary Materials}\\
\vspace{0.1in}
\vspace{0.1in}
{\huge
Functional Thermodynamics of Maxwellian Ratchets: \\ 
Randomizing and Derandomizing Behaviors, \\ 
Constructing and Deconstructing Patterns}\\[15pt]
{\large
Alexandra Jurgens and James P. Crutchfield\\
\arxiv{2003.00139}
}
\end{center}

\setcounter{equation}{0}
\setcounter{figure}{0}
\setcounter{table}{0}
\setcounter{page}{1}
\setcounter{section}{0}
\makeatletter
\renewcommand{\theequation}{S\arabic{equation}}
\renewcommand{\thefigure}{S\arabic{figure}}
\renewcommand{\thetable}{S\arabic{table}}

\section{Stochastic Processes}
\label{app:processes}

Several of the tools used here come from the theory of classical stochastic
processes, we introduce several definitions and notation for the reader less
familiar with it. A classical stochastic process $\bf{\MSym}$ is a series of
random variables and a specification of the probabilities of their
realizations. The random variables corresponding to the behaviors are denoted
by capital letters $\ldots \MSym_{t-2}, \MSym_{t-1}, \MSym_t, \MSym_{t+1},
\MSym_{t+2} \ldots$. Their realizations are denoted by lowercase letters
$\ldots \msym_{t-2}, \msym_{t-1}, \msym_t, \msym_{t+1}, \msym_{t+2} \ldots$,
with $\msym_t$ values drawn from a discrete alphabet $\Abet$, in the present
setting. Blocks are denoted as: $\MSym_{t: t+l} = \MSym_t, \MSym_{t+1}, \ldots
\MSym_{t+l-1}$, the left index is inclusive and the right one exclusive.

For our purposes, we consider \emph{stationary} stochastic processes, in which
the probability of observing behaviors is time-translation invariant:
\begin{align*}
   \Prob(X_{t:t+\ell} = x_{t:t+\ell}) = \Prob(X_{0:\ell} = x_{0:\ell}) ~,
\end{align*}
for all $t$ and $\ell$. We consider processes that have finite or infinite
Markov order and that can be generated by either finite or infinite hidden
Markov models.

\subsection{Hidden Markov Models and Unifilarity}
\label{subsec:hmmuni}

A \emph{hidden Markov model} (HMM) is a quadruple $(\SSet, \Abet, \{T^\msym\},\pi)$ consisting of:
\begin{itemize}
\setlength{\topsep}{0pt}
\setlength{\itemsep}{0pt}
\setlength{\parsep}{0pt}
\setlength{\labelwidth}{5pt}
\item $\SSet$ is the set of hidden states.
\item $\Abet$ the alphabet of symbols that the HMM emits on state-to-state
	transitions at each time step.
\item $\{T^\msym: \msym \in \Abet\}$  is the set of labeled transition matrices
	such that $T^\msym_{ij} = \Prob(\msym, \cs_j|\cs_i)$ with $\cs_i, \cs_j \in
    \SSet$. That is, $T^\msym_{ij}$ denotes the probability of the HMM
    transitioning from state $\cs_i$ to state $\cs_j$ while emitting symbol
    $\msym$.
\item $\pi$ is the stationary state distribution determined from the left
    eigenvector of $T = \sum_{\msym \in \Abet} T^\msym$ normalized in
    probability.
\end{itemize}

An HMM property that proves to be essential is unifilarity. An HMM is
\emph{unifilar} if, from each hidden state, the emitted symbol $\msym \in \Abet$
uniquely identifies the next state. Equivalently, for each labeled transition
matrix $T^\msym$, there is at most one nonzero entry in each row. Unifilarity
ensures that a sequence of emitted symbols has a one-to-finite correspondence
with sequences of hidden-state paths.

In contrast, if an HMM is nonunifilar the set of allowed hidden-state state
paths corresponding to a sequence of emitted symbols grows exponentially with
sequence length.  Nonunifilarity makes inferring the underlying states and
transitions directly from the generated output process a
computationally-challenging task.

\subsection{Measures of Complexity}
\label{app:moc}

We consider two complexity measures that have clear operational meanings: a
process' intrinsic randomness and the minimal memory resources required to
predict its behavior accurately.

The intrinsic randomness of a classical stochastic process $\bf{\MSym}$ is
measured by its \emph{entropy rate} \cite{Cove91a}:
\begin{align}
  \hmu = \lim\limits_{\ell \rightarrow \infty} \frac{H(\ell)}{\ell} ~,
  \label{eq:hmu1}
\end{align}
where $H(\ell) = \H[\Prob(\MS{0}{\ell-1})]$ is the Shannon entropy for
length-$\ell$ blocks. That is, a process' intrinsic randomness is the
asymptotic average Shannon entropy per emitted symbol---a process' entropy
growth rate.

Shannon showed that this is the same as the asymptotic value of the entropy of
the next symbol conditioned on the past \cite{Shan48a}:
\begin{align}
  \hmu = \lim\limits_{\ell \rightarrow \infty} \H[\MSym_0 | \MS{-\ell}{0}] ~.
\label{eq:hmu2}
\end{align}
This can be interpreted as how much information is gained per measurement once
all the possible structure in the sequence has been captured.

Determining $\hmu$ is possible only for a small subset of stochastic processes.
Shannon \cite{Shan48a} gave closed-form expressions for processes generated by
Markov Chains (MC), which are ``unhidden'' HMMs---they emit their states as
symbols. Making use of Eq. (\ref{eq:hmu2}), he proved that for MC-generated
processes the entropy rate is simply the average uncertainty in the next state:
\begin{align}
  \hmu =  - \sum\limits_{\cs \in \CausalStateSet} \Prob(\cs)
  \sum\limits_{\cs^\prime \in \CausalStateSet} T_{\cs \cs^\prime} \log
  T_{\cs \cs^\prime} ~,
\label{eq:mc_hmu}
\end{align}
where $T$ is the MC's transition matrix and $\CausalStateSet$ its set of states.

Another special case for which the entropy rate can be exactly computed is for
processes generated by unifilar HMMs (uHMMs) \cite{Cove91a}. This class
generates an exponentially larger set of processes than possible from MCs.
Since each infinite sequence of emitted symbols corresponds to a unique
sequence of internal states, or at most a finite number, the process entropy
rate is that of the internal MC. And so, one (slightly) adapts Eq.
(\ref{eq:mc_hmu}) to calculate $\hmu$ for these processes. The expression is
presented in Eq. (\ref{eq:ShannonEntropyRateHMM}).

A process' structure is most directly analyzed by determining its minimal
predictive presentation, its $\eM$. A simple measure of structure is then given
by the number of internal (\emph{causal}) states $|\CausalStateSet|$ or by the
\emph{statistical complexity} $\Cmu$ defined in Eq. (\ref{eq:Cmu}), which is
the Shannon information $\H[\CausalStateSet]$ stored in the causal states.
Since the set of causal states is minimal, $\Cmu$ measures of how much memory
about the past a process remembers. Said differently, $\Cmu$ quantifies the
minimum amount of memory necessary to optimally predict the process' future.

However, for processes generated by nonunifilar HMMs, both $\hmu$ and $\Cmu$
given by Eqs. (\ref{eq:ShannonEntropyRateHMM}) and (\ref{eq:Cmu}) are
incorrect. The former overestimates the generated process' $\hmu$, since
uncertainty in the next symbol is not in direct correspondence with the
uncertainty in the next internal state. In fact, there is no exact general
method to compute the entropy rate of a process generated by a generic
nonunifilar HMM. One has only the formal expression of Eq.
(\ref{eq:BlackwellHmuIntegral}) which refers to a abstract measure that, until
now, was not constructively determined. For related reasons, the statistical
complexity $\Cmu$ given by Eq. (\ref{eq:Cmu}) applied to that abstract measure
is useless---it simply diverges.

For processes generated by nonunifilar HMMs one can take a very pragmatic
approach to estimate randomness and structure from process realizations
(measured or simulated time series) using information measures for sequences of
finite-length ($\ell$), such as reviewed in Refs. \cite{Crut01a,Jame11a}. This
approximates the sequence statistics as an order-$\ell$ Markov process. The
associated conditional distributions capture only finite-range correlations:
$\Pr(\MeasSymbol_{t:\infty} | \meassymbol_{-\infty:t}) = \prod_{i=t}^{\infty}
\Pr(\MeasSymbol_i | \MeasSymbol_{i-\ell} \ldots \MeasSymbol_{i-1})$. This
approach is data-intensive and the complexity estimators have poor convergence.

Addressing the shortcomings for processes generated by nonunifilar HMMs
requires introducing the fundamental concepts of predictive features and a
process' mixed-state presentation.

\subsection{Calculating Mixed States}
\label{app:MixedStates}

The finite Markov-order approach seems to make sense empirically. However, one
would hope that, if we know the nonunfilar HMM and therefore have a model
(states and transitions) that generates the process at hand, we can calculate
randomness and structure directly from that model. One hopes to at least do
better than using slowly-converging order-$\ell$ Markov approximations. The
approach is to construct a unifilar HMM---the process' \eM---from the
nonunfilar HMM. This is done by calculating the latter's \emph{mixed states}.

Each mixed state tracks the probability distribution over the the nonunifilar
HMM's internal states, conditioned on the possible sequences of observed
symbols. In other words, the mixed states represent \emph{states of knowledge}
of the nonunifilar HMM's internal states.  This also allows one to compute the
transition dynamic between mixed states, forming a unifilar model for the same
process as generated by the original nonunifilar HMM.

Explicitly, assume that an observer has an HMM presentation $M$ for a process
$\Process$ and, before making any observations, has probabilistic knowledge of
the current state---the state distribution  $\mxst_0 = \Prob(\CausalState)$.
Typically, prior to observing any system output the best guess is $\mxst_0 =
\pi$.

Once $M$ generates a length-$\ell$ word $w = \msym_{0} \msym_{1} \ldots
\msym_{\ell-1}$ the observer's \emph{state of knowledge} of $M$'s current state
can be updated to $\mxst(w)$, that is:
\begin{align}
  \mxst_\cs(w) & \equiv \Prob(\CausalState_\ell = \cs | \MS{0}{\ell}=w, \CausalState_0 \sim \pi) ~.
\label{eq:MixedState}
\end{align}
The collection of possible \emph{states of knowledge} $\mxst(w)$ form the set
$\MxSSet$ of $M$'s \emph{mixed states}:
\begin{align*}
  \MxSSet = \{ \mxst(w): w \in \MeasAlphabet^+, \Pr(w) > 0 \} ~.
\end{align*}
And, we have the mixed-state (\emph{Blackwell}) measure
$\MxSMeasure(\mxst)$---the probability of being in a mixed state:
\begin{align*}
  \Pr(\mxst(w)) & = \Prob(\CausalState_\ell | \MS{0}{\ell}=w, \CausalState_0 \sim \pi) \Pr(w) ~.
\end{align*}

From this follows the probability of transitioning from $\mxst(w)$ to
$\mxst(w\msym)$ on observing symbol $\msym$:
\begin{align*}
  \Pr(\mxst(w\msym) | \mxst(w)) & = \Pr(\msym|\CausalState_\ell \sim \mxst(w)) ~.
\end{align*}
This defines the mixed-state dynamic $\MxSDyn$ over the mixed states.
Together the mixed states and their dynamic give the HMM's \emph{mixed-state
presentation} (MSP) $\MSP = \{\MxSSet, \MxSDyn \}$ \cite{Blac57b}.

Given an HMM presentation, though,  we can explicitly calculate its MSP. The
probability of generating symbol $\msym$ when in mixed state $\mxst$ is:
\begin{align}
  \Pr(\msym |\mxst) = \mxst \cdot T^{(\msym)} \cdot \One  ~,
\label{eq:SymbolFromMixedState}
\end{align}
with $\One$ a column vector of $1$s. Upon seeing symbol $\msym$, the current
mixed state $\mxst_t$ is updated:
\begin{align}
  \mxst_{t+1}(x) = \frac{\mxst_t \cdot T^{(\msym)}}{ \mxst_t \cdot T^{(\msym)} \cdot \One }   ~,
  \label{eq:MxStUpdate}
\end{align}
with $\mxst_0 = \mxst(\lambda) = \pi$ and $\lambda$ the null sequence.

Thus, given an HMM presentation we can calculate the mixed state of Eq. (\ref{eq:MixedState}) via:
\begin{align*}
  \frac{\pi \cdot T^{(w)}}{\pi \cdot T^{(w)} \cdot \One}   ~.
\end{align*}
The mixed-state transition dynamic is then:
\begin{align*}
  \Pr(\mxst_{t+1},\msym|\mxst_t) & = \Pr(\msym|\mxst_t) \\ & = \mxst_t \cdot T^{(\msym)} \cdot \One ~,
\end{align*}
since Eq. (\ref{eq:MxStUpdate}) tells us that, by construction, the MSP is
unifilar. That is, the next mixed state is a function of the previous and the
emitted (observed) symbol.

Transient mixed states are those state distributions after having seen
finite-$\ell$ sequences $w$, while recurrent mixed states are those remaining
with positive probability in the limit that $\ell \to \infty$. When their set
is minimized, recurrent mixed states exactly correspond to causal states
$\CausalStateSet$ \cite{Crut08b}.

Now, with a unifilar presentation one is tempted to directly apply Eqs.
(\ref{eq:ShannonEntropyRateHMM}) and (\ref{eq:Cmu}) to compute measures of
randomness and structure, but another challenge prevents this. With a small
number of exceptions, the MSP of a process generated by a nonunifilar HMM has
an uncountable infinity of states $\mxst$ \cite{Marz17a}. Practically, this
means that one cannot construct the full MSP, that direct application of Eq.
(\ref{eq:ShannonEntropyRateHMM}) to compute the entropy rate is not feasible,
and that $|\CausalStateSet|$ diverges and, typically, so does $\Cmu$.

\subsection{Entropy Rate of Nonunifilar Processes}
\label{app:HmuBlackwell}

Fortunately, when working with ergodic processes, such as those addressed here,
one can accurately estimate the MSP by generating a word $w_\ell$ of
sufficiently long length \cite{Jurg19a}. The main text addresses in some detail
how to use this to circumvent the complications of uncountable mixed states
when computing the entropy rate. Specifically, with the mixed states in hand
computationally, accurate numerical estimation of the entropy rate of a process
generated by a nonunifilar HMM is given by using the temporal average specified
in Eq. (\ref{eq:BlackwellHmuSum}). The development of that expression is given
in Ref. \cite{Jurg19a}.

This handily addresses accurately estimating the entropy rate of nonunifilar
processes. And so, we are left to tackle the issue of these process' structure
with the \emph{statistical complexity dimension.} This requires a deeper
discussion.

\subsection{Statistical Complexity Dimension}
\label{app:CmuDim}

$\Cmu$ diverges for processes generated by generic HMMs, as they are typically
nonunifilar and that, in turn, leads to an uncountable infinity of mixed
states. To quantify these processes' memory resources one tracks the rate of
divergence---the \emph{statistical complexity dimension} $\CmuDim$ of the
Blackwell measure $\mu$ on $\MxSSet$:
\begin{align}
  \CmuDim = \lim_{\CoarseGrain \to 0} - \frac{\H_\CoarseGrain [\AlternateState]}{\log_2 \CoarseGrain} ~,
\end{align}
where $\H_\CoarseGrain [Q]$ is the Shannon entropy (in bits) of the
continuous-valued random variable $Q$ coarse-grained at size $\CoarseGrain$ and
$\AlternateState$ is the random variable associated with the mixed states $\eta
\in \AlternateStateSet$.

$\CmuDim$ is determined by the measured process' entropy rate
$\widehat{\hmu^B}$, as given by Eq. (\ref{eq:BlackwellHmuSum}), and the
mixed-state process' \emph{spectrum of Lyapunov characteristic exponents}
(LCEs). The latter is calculated from an HMM's labeled transition matrices,
which map the mixed states $\eta_t \in \MxSSet$ according to Eq.
(\ref{eq:MxStUpdate}). The LCE spectrum $\Lambda = \{ \lambda_1, \lambda_2,
\ldots, \lambda_N: \lambda_i \geq \lambda_{i+1} \}$ is determined by
time-averaging the contraction rates along the $N$ eigendirections of this
map's Jacobian. The statistical complexity dimension is then bounded by a
modified form of the LCE dimension \cite{Fred83a}:
\begin{align}
  \CmuDim \leq \LCEDim ~,
\end{align}
where:
\begin{align}
  \LCEDim = k - 1 + \frac{\widehat{\hmu^B} + \sum_{i=1}^k \lambda _i}{|\lambda_{k+1}|}
\label{eq:LCEDim}
\end{align}
and $k$ is the greatest index for which $\widehat{\hmu^B} + \sum_{i=1}^k
\lambda_k > 0$. Reference \cite{Jurg19a} introduces this bound for an HMM's
statistical complexity dimension, interprets the conditions required for its
proper use, and explains in fuller detail how to calculate an HMM's LCE
spectrum.

In short, the set of mixed states generated by a generic HMM is equivalent to
the Cantor set defining the attractor of a nonlinear, \emph{place-dependent
iterated function system (IFS)}. Exactly calculating dimensions---say,
$\CmuDim$---of such sets is known to be difficult. This is why here we adapt
$\LCEDim$ to iterated function systems. The estimation is conjectured to be
accurate in ``typical systems'' \cite{Kapl79a,Fred83a,Feng09}. Even so, in
certain cases where the IFS does not meet the \emph{open set condition}
\cite{Feng09}---the relationship becomes an inequality: $\CmuDim < \LCEDim$.
This case, which is easily detected from an HMM's form, is discussed in more
detail in Ref. \cite{Jurg19a}.

\section{Mandal-Jarzynski Ratchet}
\label{app:MandalJarzynski}

To work with the Mandal-Jarzynski ratchet, we reformulated it in computational
mechanics terms, which is explained in \cref{sec:InfoEngines}. In its original
conception, the model was imagined as a single symbol (``bit'') interacting
with a dial that may smoothly transition between three positions, as shown on
the left in \cref{fig:MJ_composition}. This results in six possible states of
the joint dial-symbol system, $\{A\otimes0, A\otimes1, B\otimes0, B\otimes1,
C\otimes0, C\otimes1\}$. The transitions among these six states are modeled as
a Poisson process, where $\mathcal{R}_{ij}$ is the infinitesimal transition
probability from state $j$ to state $i$, with $i; j \in \{A\times0, \dots,
C\times1\}$ \cite{Mand012a}. The \emph{weight parameter} $\WeightParameter$, so
named because it is intended to model the effect of attaching a mass to the
side of the dial, impacts the probability of transitions among the six states
by making $0 \to 1$ transitions energetically distinct from $1 \to 0$
transitions. This creates a preferred ``rotational direction'', since bit flips
in one direction will be more energetically beneficial than the other. This is
what allows the ratchet to do useful work.

Explicitly, the transition rate matrix $\mathcal{R}$ is:
\begin{align}
  \mathcal{R} = 
  \begin{pmatrix}
    -1 & 1 & 0 & 0 & 0 & 0 \\
    1 & -2 & 1 & 0 & 0 & 0 \\
    0 & 1 & -2+\WeightParameter & 1+\WeightParameter & 0 & 0 \\
    0 & 0 & 1-\WeightParameter & -2-\WeightParameter & 1 & 0 \\
    0 & 0 & 0 & 1 & -2 & 1 \\
    0 & 0 & 0 & 0 & 1 & -1
  \end{pmatrix} ~. 
\end{align}
To express the ratchet's evolution over a single interaction interval of
length $\tau$, we calculate $\mathcal{T} (\tau, \WeightParameter) =
\left(e^{\mathcal{R}(\WeightParameter) \tau} \right)^{\intercal}$, the
transition matrix of the six-state Markov model representing the
Mandal-Jarzynski model. In turn, this six-state model with the states
$\{A\otimes0, A\otimes1, B\otimes0, B\otimes1, C\otimes0, C\otimes1\}$ may be
transformed into a three-state transducer, with states $\{ A, B, C\}$ and input
and output symbols in $\{ 0, 1\}$. To do this, we define the projection
matrices:
\begin{align*}
\mathbb{P}_0 = \begin{pmatrix}
	\mathbb{I}_3 \\
	\mathbf{0}_3 \end{pmatrix}
	~\text{and}~
  \mathbb{P}_1 = \begin{pmatrix}
  	\mathbf{0}_3 \\
	\mathbb{I}_3 \end{pmatrix}
~.
\end{align*}
Then, the transducer input-output matrices
$K^{\text{in}, \text{out}}(\tau, \WeightParameter)$ for a given $\mathcal{T}(\tau,
\WeightParameter)$ are given by:
\begin{align*}
K^{\text{in}, \text{out}}(\tau, \WeightParameter)
  = \left( \mathbb{P}_{\text{in}} \right)^{\intercal}
  \mathcal{T}(\tau, \WeightParameter) \mathbb{P}_{\text{out}}
  ~.
\end{align*}
For all $\WeightParameter \in \left(0, 1\right)$ and $\tau \in \left(0, \infty
\right]$, all four $K^{\text{in}, \text{out}}(\tau, \WeightParameter)$ are
positive definite matrices. This is what is meant by ``fully-connected, highly
stochastic'' controller---all transitions on all combinations of symbols have
positive probability. Explicitly, the probability function $f(\ldots)$
referenced in both \cref{fig:MJ_comp_table} and \cref{fig:MJ_composition} is
given by:
\begin{align}
  \Pr(\TapeVar' = \TapeSym', \TapeVar = \TapeSym,
  \RatchetStateVar_N = \RatchetStateSym,
  \RatchetStateVar_{N+1} = \RatchetStateSym')
  & = f(\TapeSym, \TapeSym', \RatchetStateSym, \RatchetStateSym', \WeightParameter, \tau)
  \nonumber \\
  & = \left( \mathbb{P}_{\TapeSym} \right)^{\intercal}
  \mathcal{T}_{\RatchetStateVar, \RatchetStateVar'}(\tau, \WeightParameter)  
  \mathbb{P}_{\TapeSym'}
  ~.
\label{eq:ProbabilityFunction}
\end{align}

\subsection{Composing a Ratchet with an Input Process' Machine}
\label{app:CompositionMJ}

Given an input process generated by an HMM with transition matrices $T^{(\TapeSym)}$, such
that $\TapeSym \in \{0, 1\}$, we may exactly calculate the transition matrices
$T'^{(\TapeSym')}$ of the output HMM:
\begin{align*}
  T'^{(\TapeSym')}_{\RatchetStateVar_N \times \InputStateVar_N,
  \RatchetStateVar_{N+1} \times \InputStateVar_{N+1}}
  = \sum_{\TapeSym}
  K^{\TapeSym, \TapeSym'}_{\RatchetStateVar_N, \RatchetStateVar_{N+1}}
  T^{(\TapeSym)}_{\InputStateVar_N, \InputStateVar_{N+1}}
  ~, 
\end{align*}
noting that the state space of the output HMM is the Cartesian product of the
state space of the transducer $\RatchetStateAlphabet$ and the state space of
the input machine $\InputStateSet$. Although presenting this in the setting of
the Mandal-Jarzynski ratchet specifically, this method applies for any input
machine and transducer, given that the transducer is able to recognize the
input \cite{Barn13a}.

That said, there are several interesting points specific to the
Mandal-Jarzynski ratchet we should highlight. As noted in the previous section,
the Mandal-Jarzynski transducer matrices are positive definite, guaranteeing
that the output machine will be nonunifilar, although disallowed state
transitions in input machines are preserved in the output. (Composing the
Mandal-Jarzynski ratchet with the Golden Mean Process in
\cref{fig:MJ_composition} illustrates this effect.) This is characteristic of
any transducer defined via the rate transition matrix method outlined above.
The conclusion is that the techniques required to analyze nonunifilar HMMs are
required in general.

\subsection{Biased Coin Parameter Sweep}
\label{app:MJ_BiasedCoin}

\begin{figure}
\centering
\includegraphics[width=\textwidth]{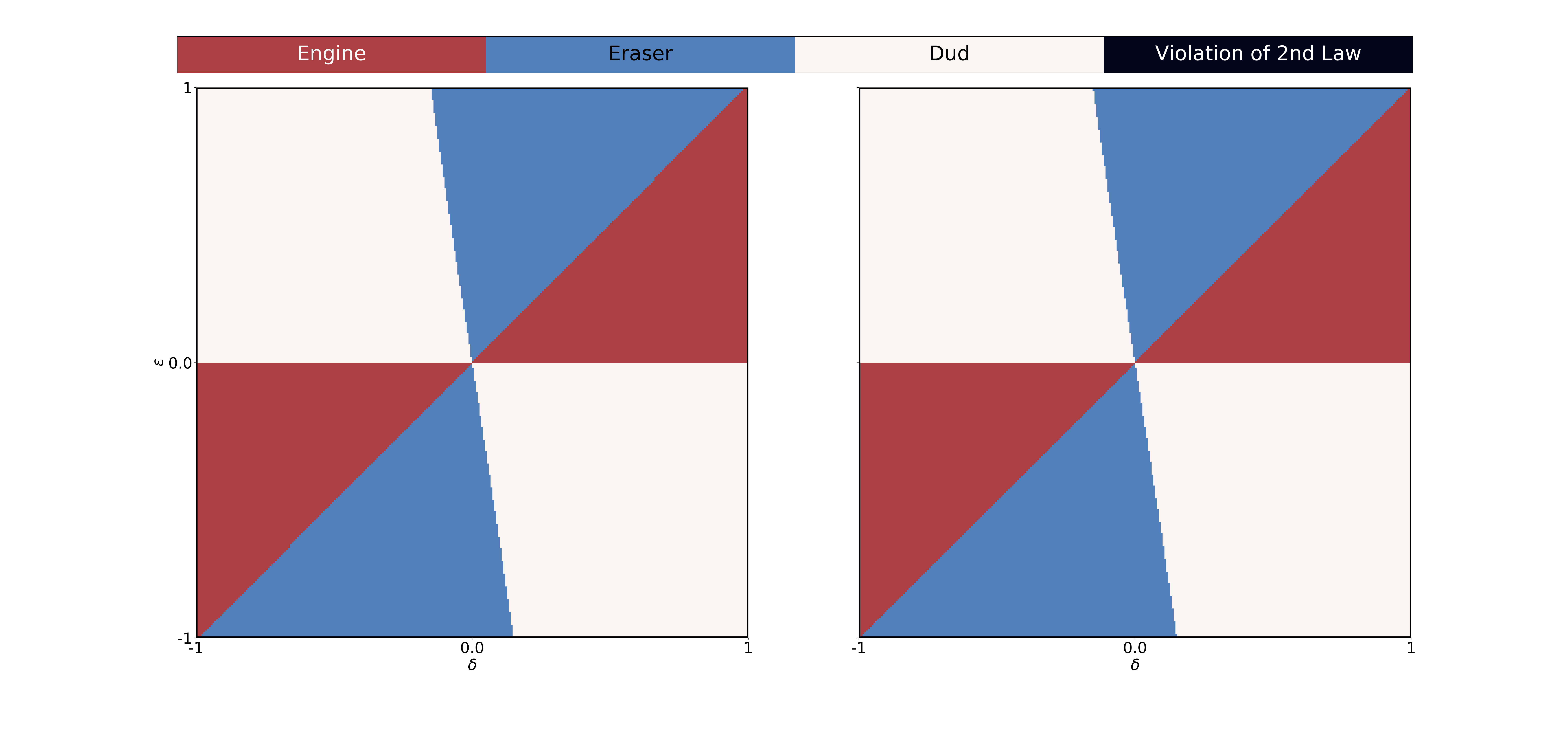}
\caption{MJ ratchet functional thermodynamic regions over $\WeightParameter \in
	[1,-1]$ and $b \in [0,10$ with $\tau = 1$. (Left) Purported functionality
	identified via by single-symbol entropy bound \cref{eq:SSBound}. (Right)
	Correct functionality identified via the entropy-rate bound IPSL
	\cref{eq:IPSL}.}
\label{fig:MJ_tau_1}
\end{figure}

As \Cref{sec:MemorylessInput} discusses, we recreated the results from Mandal
and Jarzynski's original ratchet \cite{Mand012a} using the techniques outlined
in this section and in \cref{app:processes}. There, the ratchet is driven by a
memoryless Biased Coin and the functional thermodynamic regions are
identified via \cref{eq:SSBound} \cite{Mand012a}. These results are shown in
\cref{fig:MJ_tau_1}, on the left, and demonstrate close agreement with the
original results. As previously noted, calculating the thermodynamic regions
via \cref{eq:IPSL} did not significantly change the identified regions, as can
be seen by comparison to the figure on the right. Although not shown here, we
also recreated the results at $\tau=10$, which again show strong agreement with
results reported in Ref. \cite{Mand012a}.

\subsection{Information Ratchet Mixed-State Attractor Survey}
\label{app:Grid}

\begin{figure*}
\centering
\includegraphics[width=\textwidth]{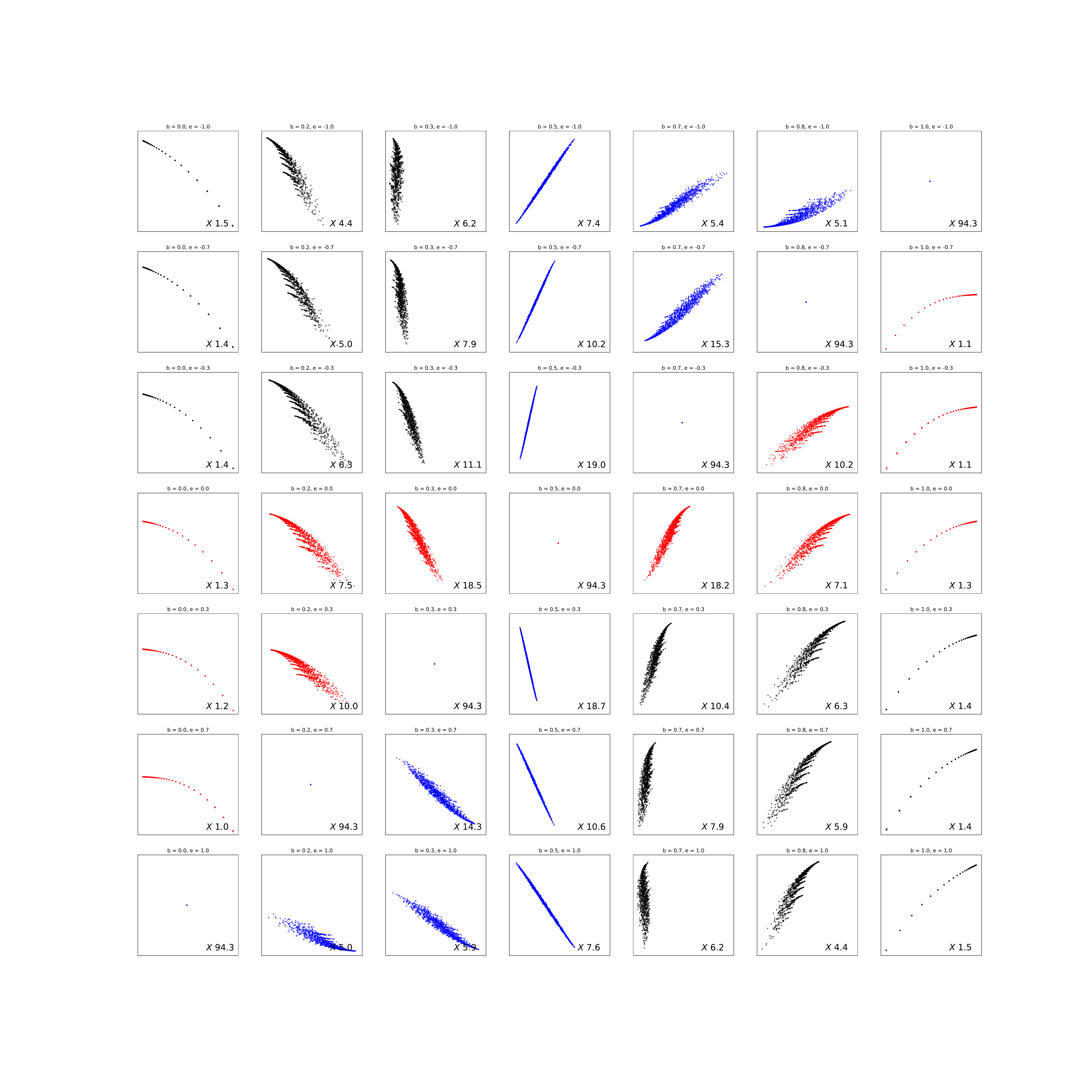}
\caption{Mixed-state attractors of the output HMMs of the Mandal-Jarzynski
	ratchet driven by a Biased Coin as a function of $\WeightParameter$ and input
	bias $\delta$, given above each square; $(\WeightParameter,\delta) \in [0,1]
	\times [-1,1]$. Each plot shows $1,000$ mixed states from the attractor at the
	magnification noted in the lower right corner. The attractors are color coded
	according to their thermodynamic functionality, with red being engines, blue
	representing erasers, and black for duds.}
\label{fig:att_grid}
\end{figure*}

To emphasize how exploring a ratchet's mixed states elucidates the underlying
physics, \cref{fig:att_grid} presents the attractors of the Mandal-Jarzynski
ratchet driven by the Biased Coin, as a function of $\WeightParameter$ and $b$,
in analogy with \cref{fig:MJ_tau_1}. Each square in the grid shows the
mixed-state attractor for the output HMM produced by the composition of the
Mandal-Jarzynski ratchet at the given $\WeightParameter$ with a Biased Coin at
the given bias $\delta$. The grid is laid out identically to the functional
thermodynamic plots above, with $\WeightParameter$ varying on the $y$-axis and
the input bias $b$ varying on the $x$-axis. Note that the squares are not at the
same scale: each is magnified to show the structure of the attractor; the
magnification factor is given in the lower right corner. Compare to
\cref{fig:MJ_simplex} to see the mixed-state attractors in further detail.
Additionally, the attractors are color coded to show thermodynamic
functionality: red for engines, blue for erasers, and black for duds.

The symmetry of the Mandal-Jarzynski ratchet around $\WeightParameter = 0$ is
revealed by how structure of the output HMM attractors is reflected and reversed
over the $\WeightParameter=\delta$ line. Along this diagonal, we see that the
mixed-state attractor collapses to a single state---a single point. This
reflects the fact that at any $\WeightParameter = \delta$ the output HMM is the
input Biased Coin, so $\langle W \rangle = \Delta H = \Delta \hmu = 0$.
Furthermore, we see that the structure of the mixed-state attractor does not
have a strong effect on the thermodynamic functionality---very similar
attractors act as duds and as erasers on each side of the $\WeightParameter=0$
line. This is as expected since, although thermodynamic functionality appears to
change suddenly, the grids in \cref{fig:MJ_GM_tau_1,fig:MJ_tau_1,fig:att_grid}
actually sweep over output machines with smoothly changing transition
probabilities. And, changes in functionality represented by the boundaries of
thermodynamic regions are actually due to small, smooth changes in the
comparative magnitude of $\langle W \rangle$ and $\Delta \hmu$.
\Cref{fig:att_grid} illustrates this clearly, as the mixed-state attractor
changes smoothly under the parameter sweep.

Note that the construction of \cref{fig:att_grid} was only possible due to the
new dynamical-systems techniques outlined in
\cref{app:MixedStates,app:HmuBlackwell}. The recently developed guarantee of
ergodicity and quick generation of mixed states allows us to easily plot and
investigate the mixed-state attractors of arbitrary HMMs. And, this allows for
parameter sweeps of attractors of HMM families and rapid calculation of their
entropy rates. The latter was required to determine the thermodynamic
functionality color coding in \cref{fig:att_grid}.

\subsection{Pattern Deconstruction and Thermodynamic Functionality}
\label{app:PatternDeconstruction}

\begin{figure}
  \centering
  \includegraphics[width=0.5\textwidth]{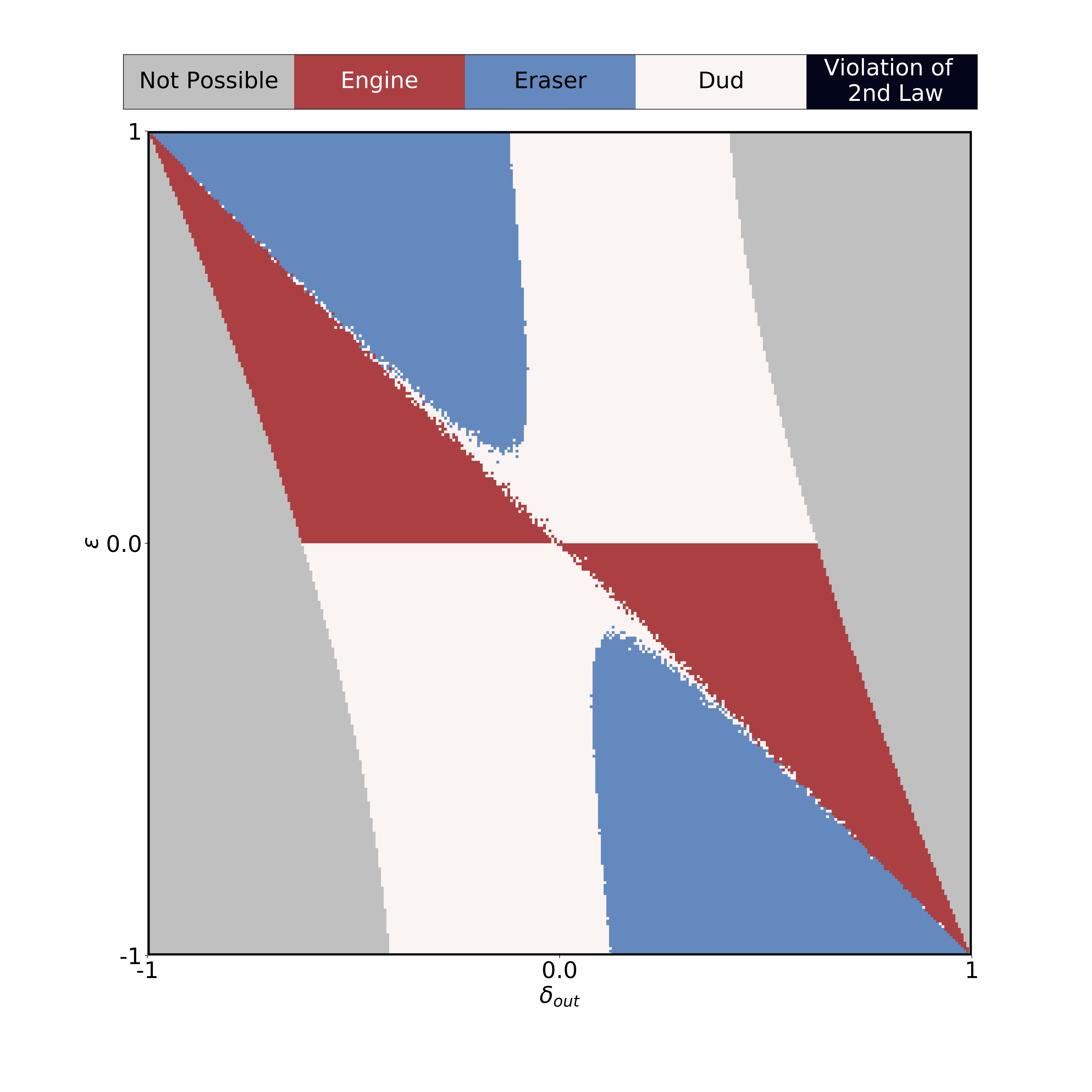}
  \caption{Functional thermodynamic regions of a ratchet pattern deconstructor with an
  interaction interval of $\tau = 0.75$. The $x$-axis sweeps over the output bias
  $\delta_{out}$, while the $y$-axis sweeps over $\WeightParameter$. As indicated,
  there are two parameter regimes where the ratchet is unable to act as a pattern
  deconstructor. In these regions, the desired output bias is not reachable by the
  machine at the given $\WeightParameter$. As $\tau \to \infty$, this region
  grows, until it encompasses every parameter combination other than
  $\WeightParameter = \delta_{out}$, which is always reachable with an input
  Biased Coin with bias $\delta_{in} = \delta_{out} = \WeightParameter$.}
  \label{fig:patt_ext}
\end{figure}

While it is relatively simple to run the Mandal-Jarzynski ratchet as an ideal
pattern constructor, forcing the ratchet to perfectly deconstruct patterns is a
more difficult task. As previously discussed in Ref. \cite{Boyd17a}, in order to
deconstruct patterns, the input and ratchet must remain synchronized. However,
the Mandel-Jarzynski ratchet, being highly stochastic, resists synchronization.
Since the input process cannot stay synced to the states of the Mandal-Jarzynski
ratchet, it must synchronize to the mixed states instead. To design an input
sequence that the ratchet transduces to an IID output process with bias $\delta$
we calculate as follows:
  \begin{enumerate}
  \setlength{\topsep}{0mm}
  \setlength{\itemsep}{-1mm}
  \setlength{\parsep}{0pt}
  \setlength{\labelwidth}{5pt}
  \item Pick a ratchet mixed state;
  \item Determine the input-output probability distribution;
  \item Calculate the input probability distribution such that
    $\Pr(\TapeVar' = 0) - \Pr(\TapeVar' = 1) = \delta$.
  \item Step forward, record the input.
  \item Use the input to update the ratchet mixed state; and
  \item Repeat the procedure starting at Step (1), using the new mixed state.
\end{enumerate}
Note that one must ensure that the output probability distribution remains
constant at each time step.

As might be suspected from the algorithm, this is not possible at all
parameters. For example, the ratchet may be so heavily biased to flip $0 \to 1$
that emitting a sequence of mainly $1$s is mathematically impossible. This is
expressed in the algorithm by finding a required input probability distribution with
a negative component. This is illustrated in \cref{fig:patt_ext}, where the
functional thermodynamic regions associated with the Mandal-Jarzynski ratchet
acting as a pattern deconstructor are shown. There are two inaccessible
regions, where the desired output is not possible for the ratchet at the given
value of $\epsilon$. As $\tau \to \infty$ these regions grow in size, until at
large $\tau$ the only parameter region capable of pattern deconstruction is
$\delta = \epsilon$. This is where the ratchet becomes memoryless, so it is
trivially a pattern deconstructor along this line.


\begin{thebibliography}{10}

\bibitem{Maxw88a}
J.~C. Maxwell.
\newblock {\em Theory of Heat}.
\newblock Longmans, Green and Co., London, United Kingdom, ninth edition, 1888.

\bibitem{Maruyama09}
K.~Maruyama, F.~Nori, and V.~Vedral.
\newblock Colloquium: The physics of {Maxwell's} demon and information.
\newblock {\em Rev. Mod. Phys.}, 81, 2009.

\bibitem{Penrose70}
O.~Penrose.
\newblock {\em Foundations of Statistical Mechanics; A Deductive Treatment}.
\newblock Oxford: Pergamon, 1970.

\bibitem{Bennet82}
C.~H. Bennet.
\newblock Thermodynamics of computation—-a review.
\newblock {\em Int. J. Theor. Phys.}, 21, 1982.

\bibitem{Land61a}
R.~Landauer.
\newblock Irreversibility and heat generation in the computing process.
\newblock {\em IBM J. Res. Develop.}, 5(3):183--191, 1961.

\bibitem{Mand012a}
D.~Mandal and C.~Jarzynski.
\newblock Work and information processing in a solvable model of {Maxwell's}
  demon.
\newblock {\em Proc. Natl. Acad. Sci. USA}, 109(29):11641--11645, 2012.

\bibitem{Boyd15a}
A.~B. Boyd, D.~Mandal, and J.~P. Crutchfield.
\newblock Identifying functional thermodynamics in autonomous {Maxwellian}
  ratchets.
\newblock {\em New J. Physics}, 18:023049, 2016.

\bibitem{Boyd16c}
A.~B. Boyd, D.~Mandal, and J.~P. Crutchfield.
\newblock Correlation-powered information engines and the thermodynamics of
  self-correction.
\newblock {\em Phys. Rev. E}, 95(1):012152, 2017.

\bibitem{Boyd16e}
A.~B. Boyd, D.~Mandal, P.~M. Riechers, and J.~P. Crutchfield.
\newblock Transient dissipation and structural costs of physical information
  transduction.
\newblock {\em Phys. Rev. Lett.}, 118:220602, 2017.

\bibitem{Boyd17a}
A.~B. Boyd, D.~Mandal, and J.~P. Crutchfield.
\newblock Thermodynamics of modularity: Structural costs beyond the {Landauer}
  bound.
\newblock {\em Phys. Rev. X}, 8(3):031036, 2018.

\bibitem{Cove91a}
T.~M. Cover and J.~A. Thomas.
\newblock {\em Elements of Information Theory}.
\newblock Wiley-Interscience, New York, 1991.

\bibitem{Boyd16d}
A.~B. Boyd, D.~Mandal, and J.~P. Crutchfield.
\newblock Leveraging environmental correlations: The thermodynamics of
  requisite variety.
\newblock {\em J. Stat. Phys.}, 167(6):1555--1585, 2016.

\bibitem{Blac57b}
D.~Blackwell.
\newblock The entropy of functions of finite-state {Markov} chains.
\newblock In {\em Transactions of the first Prague conference on information
  theory, Statistical decision functions, Random processes}, volume~28, pages
  13--20, Prague, Czechoslovakia, 1957. Publishing House of the Czechoslovak
  Academy of Sciences.

\bibitem{Jurg19a}
A.~Jurgens and J.~P. Crutchfield.
\newblock Shannon entropy rate and statistical complexity dimension of hidden
  {Markov} processes.
\newblock {\em in preparation}, 2020.

\bibitem{Crut12a}
J.~P. Crutchfield.
\newblock Between order and chaos.
\newblock {\em Nature Physics}, 8(January):17--24, 2012.

\bibitem{Crut13a}
J.~P. Crutchfield, P.~Riechers, and C.~J. Ellison.
\newblock Exact complexity: {Spectral} decomposition of intrinsic computation.
\newblock {\em Phys. Lett. A}, 380(9-10):998--1002, 2016.

\bibitem{Marz17a}
S.~E. Marzen and J.~P. Crutchfield.
\newblock Nearly maximally predictive features and their dimensions.
\newblock {\em Phys. Rev. E}, 95(5):051301(R), 2017.

\bibitem{Kish2012}
L.B. Kish and C.~G. Granqvist.
\newblock Energy requirement of control: Comments on {Szilard's} engine and
  {Maxwell's} demon.
\newblock {\em EuroPhys. Lett.}, 98, 2012.

\bibitem{Sagawa12}
T.~Sagawa and M.~Ueda.
\newblock Fluctuation theorem with information exchange: Role of correlations
  in stochastic thermodynamics.
\newblock {\em Phys. Rev. Lett.}, 109, 2012.

\bibitem{Kundu12}
A.~Kundu.
\newblock Nonequilibrium fluctuation theorem for systems under discrete and
  continuous feedback control.
\newblock {\em Phys. Rev. E}, 86, 2012.

\bibitem{Abreu12}
A.~Abreu and U.~Seifert.
\newblock Thermodynamics of genuine nonequilibrium states under feedback
  control.
\newblock {\em Phys. Rev. Lett.}, 108, 2012.

\bibitem{Vaikuntanathan11}
S.~Vaikuntanathan and C.~Jarzynski.
\newblock Modeling {Maxwell's} demon with a microcanonical {Szilard} engine.
\newblock {\em Phys. Rev. E}, 83, 2011.

\bibitem{Abreu11}
D.~Abreu and U.~Seifert.
\newblock Extracting work from a single heat bath through feedback.
\newblock {\em EuroPhys. Lett.}, 94, 2011.

\bibitem{Granger11}
L.~Granger and H.~Krantz.
\newblock Thermodynamic cost of measurements.
\newblock {\em Phys. Rev. E}, 84, 2011.

\bibitem{Horowitz11}
J.~M. Horowitz and J.~M.~R. Parrondo.
\newblock Thermodynamic reversibility in feedback processes.
\newblock {\em EuroPhys. Lett.}, 95, 2011.

\bibitem{Ponmurugan11}
M.~Ponmurugan.
\newblock Generalized detailed fluctuation theorem under nonequilibrium
  feedback control.
\newblock {\em Phys. Rev. E}, 82, 2010.

\bibitem{Toya10a}
S.~Toyabe, T.~Sagawa, M.~Ueda, E.~Muneyuki, and M.~Sano.
\newblock Experimental demonstration of information-to-energy conversion and
  validation of the generalized {Jarzynski} equality.
\newblock {\em Nature Physics}, 6:988--992, 2010.

\bibitem{Sagawa10}
T.~Sagawa and M.~Ueda.
\newblock Generalized {Jarzynski} equality under nonequilibrium feedback
  control.
\newblock {\em Phys. Rev. Lett.}, 104, 2010.

\bibitem{Cao04}
F.~J. Cao, L.~Dinis, and J.~M.~R. Parrondo.
\newblock Feedback control in a collective flashing ratchet.
\newblock {\em 93}, Phys.Rev. Lett., 2004.

\bibitem{Touchette00}
H.~Touchette and S.~Lloyd.
\newblock Information-theoretic limits of control.
\newblock {\em Phys. Rev. Lett.}, 84, 2000.

\bibitem{Ito13}
S.~Ito and T.~Sagawa.
\newblock Information thermodynamics on causal networks.
\newblock {\em Phys. Rev. Lett.}, 111, 2013.

\bibitem{Hartich14}
D.~Hartich, A.~C. A.~C. Barato, and U.~Seifert.
\newblock Stochastic thermodynamics of bipartite systems: transfer entropy
  inequalities and a {Maxwell's} demon interpretation.
\newblock {\em J. Stat. Mech}, 2014.

\bibitem{Horowitz14}
J.~M. Horowitz and M.~Esposito.
\newblock Thermodynamics with continuous information flow.
\newblock {\em Phys. Rev. X}, 4, 2014.

\bibitem{Horowitz15}
J.~M. Horowitz.
\newblock Multipartite information flow for multiple {Maxwell} demons.
\newblock {\em J. Stat. Mech.}, 2015.

\bibitem{Strasberg13}
P. Strasberg, G.~Schaller, T.~Brandes, and M.~Esposito.
\newblock Thermodynamics of a physical model implementing a {Maxwell} demon.
\newblock {\em Phys. Rev. Lett.}, 110, 2013.

\bibitem{Esposito12}
M.~Esposito and G.~Schaller.
\newblock Stochastic thermodynamics for `{Maxwell} demon' feedbacks.
\newblock {\em EuroPhys. Lett.}, 99, 2012.

\bibitem{Horowitz14b}
J.~M. Horowitz and H.~Sandberg.
\newblock Second-law-like inequalities with information and their
  interpretations.
\newblock {\em New J. Phys.}, 16, 2014.

\bibitem{Barato14}
A.~C. Barato and U.~Seifert.
\newblock Unifying three perspectives on information processing in stochastic
  thermodynamics.
\newblock {\em Phys. Rev. Lett.}, 112, 2014.

\bibitem{Horowitz13}
J.~M. Horowitz, T.~Sagawa, and J.~M.~R. Parrondo.
\newblock Imitating chemical motors with optimal information motors.
\newblock {\em 2013}, 111, Phys. Rev. Lett.

\bibitem{Mandal13}
D.~Mandal, H.~T. Quan, and C.~Jarzynski.
\newblock Maxwell's refrigerator: An exactly solvable model.
\newblock {\em Phys. Rev. Lett.}, 111, 2013.

\bibitem{Barato13}
A.~C. Barato and U.~Seifert.
\newblock Anautonomous and reversible {Maxwell's} demon.
\newblock {\em Europhys. Lett.}, 101, 2013.

\bibitem{Hoppenau14}
J.~Hoppenau and A.~Engel.
\newblock On the energetics of information exchange.
\newblock {\em Europhys. Lett.}, 105, 2014.

\bibitem{Lu14}
Z.~Lu, D.~Mandal, and C.~Jarzynski.
\newblock Engineering {Maxwell's} demon.
\newblock {\em Phys. Today}, 67, 2014.

\bibitem{Um15}
J.~Um, H.~Hinrichsen, C.~Kwon, and H.~Park.
\newblock Total cost of operating an information engine.
\newblock {\em New J. Phys}, 17, 2015.

\bibitem{Cove06a}
T.~M. Cover and J.~A. Thomas.
\newblock {\em Elements of Information Theory}.
\newblock Wiley-Interscience, New York, second edition, 2006.

\bibitem{Lu14a}
Z.~Lu, D.~Mandal, and C.~Jarzynski.
\newblock Engineering {Maxwell's} demon.
\newblock {\em Physics Today}, 67(8):60--61, January 2014.

\bibitem{Cao15}
Y.~Cao, Z.~Gong, and H.~T. Quan.
\newblock Thermodynamics of information processing based on enzyme kinetics: An
  exactly solvable model of an information pump.
\newblock {\em Phys. Rev. E}, 91, 2017.

\bibitem{Shiraishi15}
N.~Shiraishi, S.~Ito, K.~Kawaguchi, and T.~Sagawa.
\newblock Role of measurement-feedback separation in autonomous {Maxwell's}
  demon.
\newblock {\em New J. Phys.}, 17, 2015.

\bibitem{Diana13}
G.~Diana, G.~B. Bagci, and M.~Esposito.
\newblock Finite-time erasing of information stored in fermionic bits.
\newblock {\em Phys. Rev. E}, 8, 2013.

\bibitem{Chap15a}
A.~Chapman and A.~Miyake.
\newblock How can an autonomous quantum {Maxwell} demon harness correlated
  information?
\newblock arXiv:1506.09207, 2015.

\bibitem{Parrondo15}
J.~M.~R. Parrondo, J.~M. Horowitz, and T.~Sagawa.
\newblock Thermodynamics of information.
\newblock {\em Nature Physics}, 11, 2015.

\bibitem{Riechers19}
P.~M. Riechers.
\newblock {\em Transforming Metastable Memories: The Nonequilibrium
  Thermodynamics of Computation}.
\newblock SFI Press, 2019.
\newblock arXiv:1808.03429.

\bibitem{Ashby60}
W.~R. Ashby.
\newblock {\em An Introduction to Cybernetics, 2nd Edn.}
\newblock Wiley, New York, 1960.

\bibitem{Ehrenberg80}
M.~Ehrenberg and C.~Blomberg.
\newblock Thermodynamic constraints on kinetic proofreading in biosynthetic
  pathways.
\newblock {\em Biophys. J.}, 31, 1980.

\bibitem{Beru12a}
A.~Berut, A.~Arakelyan, A.~Petrosyan, S.~Ciliberto, R.~Dillenschneider, and
  E.~Lutz.
\newblock Experimental verification of {Landauer's} principle linking
  information and thermodynamics.
\newblock {\em Nature}, 483:187--190, 2012.

\bibitem{Saga12a}
T.~Sagawa.
\newblock Thermodynamics of information processing in small systems.
\newblock {\em Prog. Theo. Phys.}, 127(1):1--56, 2012.

\bibitem{Seif12a}
U.~Seifert.
\newblock Stochastic thermodynamics, fluctuation theorems and molecular
  machines.
\newblock {\em Rep. Prog. Phys.}, 75:126001, 2012.

\bibitem{Cont19a}
T.~Conte et~al.
\newblock Thermodynamic computing.
\newblock {\em arxiv:1911.01968}, 2019.

\bibitem{Wims19a}
G.~Wimsatt, O.-P. Saira, A.~B. Boyd, M.~H. Matheny, S.~Han, M.~L. Roukes, and
  J.~P. Crutchfield.
\newblock Harnessing fluctuations in thermodynamic computing via time-reversal
  symmetries.
\newblock arXiv:1906.11973, 2019.

\bibitem{Sair19c}
O.-P. Saira, M.~H. Matheny, R.~Katti, W.~Fon, G.~Wimsatt, S.~Han, J.~P.
  Crutchfield, and M.~L. Roukes.
\newblock Nonequilibrium thermodynamics of erasure with superconducting flux
  logic.
\newblock {\em Phys. Rev. Res.}, 2:013249, 2020.

\bibitem{Shan48a}
C.~E. Shannon.
\newblock A mathematical theory of communication.
\newblock {\em Bell Sys. Tech. J.}, 27:379--423, 623--656, 1948.

\bibitem{Crut01a}
J.~P. Crutchfield and D.~P. Feldman.
\newblock Regularities unseen, randomness observed: Levels of entropy
  convergence.
\newblock {\em CHAOS}, 13(1):25--54, 2003.

\bibitem{Jame11a}
R.~G. James, C.~J. Ellison, and J.~P. Crutchfield.
\newblock Anatomy of a bit: {Information} in a time series observation.
\newblock {\em CHAOS}, 21(3):037109, 2011.

\bibitem{Crut08b}
C.~J. Ellison, J.~R. Mahoney, and J.~P. Crutchfield.
\newblock Prediction, retrodiction, and the amount of information stored in the
  present.
\newblock {\em J. Stat. Phys.}, 136(6):1005--1034, 2009.

\bibitem{Fred83a}
P.~Frederickson, J.~Kaplan, E~Yorke, and J.~Yorke.
\newblock The {Lyapunov} dimension of strange attractors.
\newblock {\em J. Diff. Eqs}, 49(2):185--207, 1983.

\bibitem{Kapl79a}
J.~Kaplan and J.~Yorke.
\newblock Chaotic behavior of multidimensional difference equations.
\newblock In {\em Functional Differential Equations and Approximation of Fixed
  Points}, volume 730 of {\em Lecture Notes in Mathematics}, pages 204--227.
  Springer, 1979.

\bibitem{Feng09}
D.~Feng and H.~Hu.
\newblock Dimension theory of iterated function systems.
\newblock {\em Comm. Pure App. Math.}, 62(11):1435--1500, 2009.

\bibitem{Barn13a}
N.~Barnett and J.~P. Crutchfield.
\newblock Computational mechanics of input-output processes: {Structured}
  transformations and the $\epsilon$-transducer.
\newblock {\em J. Stat. Phys.}, 161(2):404--451, 2015.

\end{thebibliography}
\end{document}